\documentclass[12pt]{iopart}

\usepackage{amssymb}
\usepackage{epstopdf}
\usepackage[english]{babel}
\usepackage[square, numbers, sort, compress]{natbib}
\usepackage{graphicx}
\usepackage{pgf}
\usepackage{dcolumn}
\usepackage{bm}
\usepackage{xcolor}
\usepackage[unicode,
colorlinks = true,
linkcolor = magenta,
urlcolor  = blue,
citecolor = blue,
anchorcolor = blue,
filecolor=blue]{hyperref}

\newcommand{\Eqref}[1]{Equation~(\ref{#1})}

\def\ie{{i.e.,~}}
\def\eg{{e.g.,~}}

\def\cA{\mathcal{A}}

\def\cO{\mathcal{O}}

\def\cS{\mathcal{S}}

\def\x{\vec{x}}

\def\dt{\Delta t}
\def\dWh{\Delta \xi^H}

\newcommand{\avg}[1]{\left\langle #1 \right \rangle}

\newcommand{\dint}[1]{{\rm d}{#1}\,}

\newcommand{\Pxi}{P_{\xi}}

\usepackage{appendix}

\begin{document}

	\title{
		Inferring nonlinear fractional diffusion processes from single trajectories
	}
	
	\author{Johannes A. Kassel$^{1, \ast}$, Benjamin Walter$^{2, \dagger}$, and Holger Kantz$^1$}%
	\address{$^1$ Max Planck Institute for the Physics of Complex Systems, N\"othnitzer Straße 38, 01187 Dresden, Germany
	}%
	\address{$^2$ Department of Mathematics, Imperial College London, 180 Queen's Gate, SW7 2AZ London, United Kingdom}
	\ead{$^{\ast}$jkassel@pks.mpg.de, $^{\dagger}$b.walter@imperial.ac.uk}

	\date{\today}
	
	\begin{abstract} 
		We present a method 
		to infer 
		the arbitrary space-dependent drift 
		and diffusion 
		of a 
		nonlinear stochastic  model 
		driven by multiplicative fractional Gaussian noise
		from a single trajectory.
		Our method, fractional Onsager-Machlup optimisation (fOMo), 
		introduces a maximum likelihood estimator 
		by minimising a field-theoretic action 
		which we construct from the observed time series.
		We successfully test fOMo  for a wide range of Hurst exponents using artificial data with strong nonlinearities,  and
		apply it to a data set of daily mean temperatures.
		We further 
		highlight
		the significant systematic estimation errors when ignoring non-Markovianity, underlining the need for nonlinear fractional inference methods when studying real-world long-range
		\mbox{(anti-)correlated} systems.
		
	\end{abstract}
	
	\noindent{\it Keywords\/}: statistical inference, fractional Brownian motion, nonequilibrium statistical mechanics, maximum likelihood estimation, single-trajectory measurements, time series analysis, anomalous diffusion
	
	\submitto{\NJP}
	

	\section{Introduction}
	The dynamic behaviour of complex systems comprising many degrees of freedom poses significant challenges to analytic description and often eludes physical intuition.
	Typically, measurements only access few slowly evolving degrees of freedom which fluctuate stochastically, indicating the presence of hidden interactions within the system.
	If the system is organised in hierarchical self-similar subsystems, their fractal nature may be mirrored in scale-free correlated, \ie \emph{fractional}, fluctuations leading to non-negligible departure from Markovianity \cite{Metzler2014}.
	Important examples of systems displaying fractional correlations include the climate \cite{Hurst1951, Bunde2001, Eichner2003, Bunde2005,  Kantelhardt2006}, DNA \cite{Voss1992, Chatzidimitriou1993}, cell motility \cite{Hofling2013,Wan2014}, and the brain \cite{Palva2013,Teka2014}.
	
	Recent efforts in studying such processes have focused on machine learning-based prediction; despite their power \cite{Munoz-Gil2021}, such approaches often lack interpretability \cite{Bzdok2018} or robustness in real-world scenarios \cite{Carleo2019, Cichos2020}.
	In order to build an analytic and intuitive understanding of such fractional processes, \emph{statistical inference}, in which parameters of a conjectured stochastic model are inferred from experimental data, may help in providing clearer explainability and interpretation. 
	Recent inference methods focused on either linear fractional \cite{Rao2011, Burnecki2014, Graves2015, Kubilius2017, Thapa2018}, or nonlinear non-fractional processes \cite{Siegert1998, Ragwitz2001, Boettcher2006, Beheiry2015, Gouasmi2017, Lehle2018, PerezGarcia2018, Tabar2019Book, Brueckner2020, Frishman2020, Ferretti2020, Willers2021}. 
	In scenarios, however, where nonlinear dynamics and fractional fluctuations appear simultaneously, their complicated interplay leads to many emergent features not present in either purely nonlinear or fractional models. Hence, if these phenomenologically richer scenarios are to be studied, more generalised statistical inference methods are required.
	In this article, we propose an estimation method for a fully nonlinear, fractional model of time series, recovering space dependent drift and diffusion parameters from a single trajectory. Thus, it is suitable for scenarios in which only single realisations are available \cite{Weiss2013,Saelens2019}.
	In developing the estimation method, we draw on the field-theoretic Onsager-Machlup formalism \cite{Onsager1953,Durr1978,Taeuber2014,Cugliandolo2017} which we generalise to arbitrarily correlated, and in particular fractional, processes.
	We benchmark the method using both synthetic and real world data, where it reconstructs nonlinear parameters and statistics with excellent accuracy.
	We further show that ignoring fractional correlations leads to systematically wrong parameter inferences.
	
	\begin{figure}[t]
		\centering
		\includegraphics[width=.7\columnwidth]{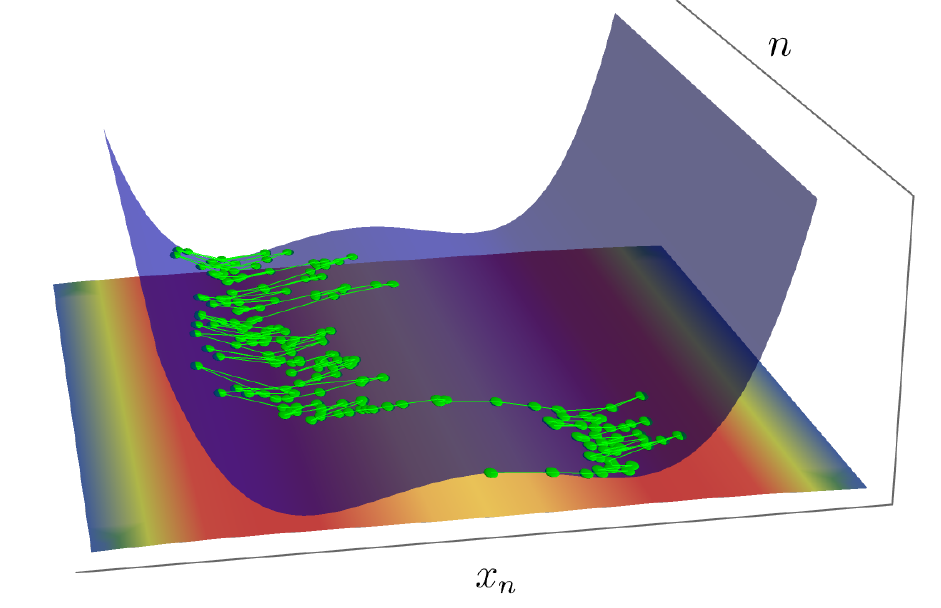}
		\caption{The stochastic difference equation given in \Eqref{eq:FLE_def} describes the stochastic evolution of a process $x_n$ subject to a nonlinear drift $f(x) = -V^{\prime}(x)$ ($V(x)$ shown as blue shade) and a non-homogeneous diffusion $g(x)$ (rainbow coloured ground). The process is driven by multiplicative fractional driving noise $g(x)\dWh$. We propose an algorithm which infers both $f(x), g(x)$ from a single finite trajectory (green dots). }
		\label{fig:setup}
	\end{figure}
	
	\section{Fractional processes}
	We discuss a stochastic model for nonlinear time series described by an \emph{It\^{o}}-type stochastic difference equation driven by fractional noise
	\begin{equation}
	x_{n+1} = x_n + f(x_n) \dt + g(x_n) \dWh_{n+1} \ ,
	\label{eq:FLE_def}
	\end{equation}
	where $f(x) , g(x) $ denote space-dependent drift and diffusion terms of dimension $[f] = [x]/[\mathrm{time}], [g]=[x]/[\mathrm{time}]^H$. The process is illustrated in Figure~\ref{fig:setup}. 
	\Eqref{eq:FLE_def} is the first-order discretisation (Euler-Maruyama scheme) of a stochastic differential equation driven by fractional noise \cite{Pipiras2017}.
	While $\dt$ denotes the time step of the discretisation, $\dWh_n$ is a discretised fractional Gaussian noise (fGN) \cite{Mandelbrot1968, Mandelbrot1969}, defined by its correlation matrix
	\begin{equation}
	\fl	\label{eq:corr_fgn}
	C_{mn} = \avg{\dWh_m \dWh_n} = (\dt)^{2H} 
	\left( \left| n-m +1 \right|^{2H} + \left| n-m - 1 \right|^{2H} - 2 \left|n - m \right|^{2H} \right) \ .
	\end{equation}
	Choosing $H = \frac{1}{2}$, the noise $\dWh_n$ is discrete standard Gaussian white noise. For $H > \frac12$ ($H < \frac12$), the increments $\dWh_n$ are positively (negatively) correlated.

	\section{Inference method}
	We propose an algorithm that given a single measured trajectory $\x = \left\{  x_0, x_1, \ldots, x_{N} \right\}$ sampled at $\Delta t$ infers the functional parameters $f(x), g(x)$ as introduced in \Eqref{eq:FLE_def}. 
	We argue that $f(x)$ and $g(x)$ only introduce short-range correlations, which exponentially decay for a stationary process. We therefore expect that the asymptotic scaling of the correlation function of \Eqref{eq:FLE_def} is solely determined by $H$ and $\Delta t$ and does not depend on $f(x), g(x)$.
	We assume that the Hurst parameter has already been  estimated previously using  established methods (
	\eg \cite{Peng1994, Hoell2019,Abry1998,Bo2019}).
	
	The central idea is to find the functions $f$ and $g$ which maximise the log-likelihood $\ln P[\vec{x}|f,g]$ of the observed trajectory $\x$. 
	The likelihood is obtained by measuring the likelihood of the noise realisation $\vec{\dWh}=\{ \dWh_1,...,\dWh_N\}$, which together with $f$ and $g$ produces the measurement $\x$.
	This realisation is found by inversion of \Eqref{eq:FLE_def}, 
	\begin{equation}
	\label{eq:emp_noise}
	\left(\vec{\dWh}\left( \x\right)\right)_n = \frac{x_{n} - x_{n-1} - f(x_{n-1}) \dt}{ g(x_{n-1})}\ .
	\end{equation}
	The probability to observe a measurement $\x$ then follows from $\Pxi[\vec{\dWh}]$, the probability distribution of a noise realisation, by a standard change of variables in $\mathbb{R}^N$,  
	\begin{equation}
	P[\vec{x}|f,g] =  \Pxi\left[\vec{\dWh}\left(\x\right)\right] \left| \frac{\partial \vec{\dWh}}{\partial \x} \right|,
	\end{equation} 
	where the nonlinear transform (Equation~\ref{eq:emp_noise}) induces a Jacobian determinant accounting for the measure change from $\prod_n \dint{\dWh_n}$ to $\prod_n \dint{x_n}$ \cite{Devore2012}.
	
	Since $\vec{\dWh}$ is discrete fractional Gaussian noise, its probability distribution is a $N$-dimensional Gaussian distribution. We write the distribution field-theoretically as
	\begin{equation}
	\Pxi = \left(\left(2 \pi \right)^N \det[C]\right)^{-\frac{1}{2}} \exp \left( - \cA[\dWh] \right)
	\end{equation}
	where $\cA$ is the \emph{free action}
	\begin{equation}
	\cA[\dWh] = \frac12 \left(\vec{\dWh} \right)^T C^{-1} \vec{\dWh}\ ,
	\end{equation}
	and $C^{-1}$ is the matrix inverse of the correlation matrix given in  \Eqref{eq:corr_fgn} \cite{Zinn-Justin2002}. Here, the non-diagonal correlation matrix accounts exactly for the full memory of the process.
	
	The determinant $\left|\partial \vec{\dWh}/\partial \x\right|$ itself is readily evaluated as ${\partial \dWh_n}/{\partial x_m}$ is a lower triangular matrix with all entries vanishing for $m>n$ (see \Eqref{eq:emp_noise}). Hence, the determinant is given by the product over the diagonal entries,
	\begin{equation}
	{\partial \dWh_n}/{\partial x_n} = g(x_{n-1})^{-1} \,.
	\end{equation}
	This results in a log-likelihood of the measurement
	\begin{equation}
	\fl    \ln P[\x|f,g]
	\label{eq:log_likelihood} 
	= -\frac{1}{2} \sum_{m,n=1}^N \frac{x_{m} - x_{m-1} - f(x_{m-1}) \dt}{ g(x_{m-1})} C^{-1}_{m,n} \frac{x_{n} - x_{n-1} - f(x_{n-1}) \dt}{ g(x_{n-1})} - \sum_{n=0}^{N-1} \ln \left|g(x_n)\right|, \quad
	\end{equation}
	where we drop added normalisation constants independent of $f$ or $g$. 
	
	In the field-theoretic literature, expressions of the form of \Eqref{eq:log_likelihood} are well known for diagonal correlation matrices $C$, corresponding to uncorrelated noise with $H=\frac{1}{2}$, where they are referred to as \emph{Onsager-Machlup} actions \cite{Onsager1953,Durr1978,Taeuber2014,Cugliandolo2017}. These have been extensively studied to characterise the ``most likely path'' of a specifically parametrised stochastic process \cite{Horsthemke1975,Durr1978,Adib2008, Kappler2020, Gladrow2021, Kieninger2021}. In this article, we take the opposite direction and use Onsager-Machlup theory to determine which parameters render the observed path the likeliest. 
	This means that we henceforth interpret the term in \Eqref{eq:log_likelihood} as a \emph{path action} $\cS[f,g| \x] = -\ln P[\x|f,g]$ in the parameters $f$ and $g$, while the path $x$ now in turn serves as a parametrisation.
	In doing so, we draw a connection from established (Markovian) maximum likelihood estimators
	\cite{Kleinhans2012, PerezGarcia2018, Ferretti2020} to the Onsager-Machlup formalism. We generalise these results
	to fractionally driven processes and hence refer to the estimation method based on maximising \Eqref{eq:log_likelihood} as \emph{fractional Onsager-Machlup optimisation} (fOMo).
	
	In order to estimate the optimal parameters $f$ and $g$ that maximise the likelihood of an observation $\x$, one needs to find the minimum of the path action $\cS[f,g| \x] = -\ln P[\x|f,g]$ for some fixed observed $\x$. In order to do so, some finite-dimensional parametrisation of $f$ and $g$ is required; by introducing a set of suitable basis functions $\chi_p(x), \chi'_{p'}(x)$, one may rewrite $f(x) = \sum_{p=1}^P f_p \chi_p(x), g(x) = \sum_{p'=1}^{P'} g_p \chi_{p'}(x)$ for some finite $P,P'$ (chosen to be much smaller than $N$). The optimisation then takes place in the finite coefficients $f_p, g_{p'}$ of the parametrisation. Suitable basis functions include polynomials or indicator functions of disjoint intervals.
	
	In finding the minima of the parametrised path action
	\begin{equation}
	\cS[\{ f_p \}, \{g_{p'} \}] = \cS\left[\sum_{p=1}^P f_p \chi(p), \sum_{p'=1}^{P'} g_p \chi_{p'}(x)\right]
	\end{equation}
	in $\{ f_p \}, \{g_{p'} \}$, it remains to discuss the uniqueness of any solution. As a derivation provided in \ref{app:convexity} shows, the action has a unique global minimum  in either the parameters of drift, $\{ f_p \}$, or diffusion, $\{g_p\}$, when keeping the respective other set of parameters constant. When, however, considering the path action with fully free parameters $\{ f_p \}, \{g_{p'} \}$, the picture is different. On the one hand, the action is smooth, globally bounded from below and remains partially convex along both $f_p, g_p$ since both $\partial^2 \cS/ (\partial_{f_p} \partial_{f_q}) > 0 , \partial^2 \cS/ (\partial_{g_p} \partial_{g_q}) > 0 $ (see \ref{app:convexity}), therefore is assured to have a minimum, and excludes the possibility of more than one isolated local minimum. On the other hand, the set of points where $\cS$ assumes its minimum could hypothetically be a submanifold, and the optimal choice of $f,g$ could be non-unique. This degeneracy, however, is rooted in a physical ambiguity: for a specific observed time series $\x$, it is mathematically possible to modify both drift and diffusion simultaneously in a way that leaves the path probability of $\x$ unchanged. Consequently, the set of possible $\{f_p, g_{p'} \}$ which are \emph{equally} compatible with an observation $\x$ may not be unique.
	
	The problem of finding the optimum of \Eqref{eq:log_likelihood} in $f$ and $g$ is generally solved numerically.
	Nonetheless, in certain cases the optimum of $\cS$ can be found analytically.
	These special cases are discussed in the following section.
	
	\section{Exact results}
	The estimation method corresponds to finding the configuration $f(x), g(x)$ that minimises $-\ln P[\x|f,g]$. Interpreting $\cS[f,g|\x] = -\ln P[\x|f,g]$ as a field theoretic action in $f(x),g(x)$ that is parametrised by the inherently stochastic measurement $\x$ and therefore random (``disordered''), fOMo amounts to finding the minimum of $\cS$, where simultaneously $\delta \cS/ \delta f \equiv \delta \cS/\delta g \equiv 0$. 
	
	If $g$ is fixed, but not necessarily constant, the minimum in $f$ can be found analytically. We introduce the \emph{empirical propagator} 
	\begin{equation}
	[G(g)]_{m,n} = \frac{C^{-1}_{m,n}}{g(x_{m-1})g(x_{n-1})} \ ,
	\end{equation}
	as well as the \emph{empirical velocity} $v_n = (x_{n}-x_{n-1})/\Delta t$.
	Inserting these into \Eqref{eq:log_likelihood}, the action  then reads
	\begin{equation}
	\label{eq:log_likelihood_f}
	\!\!\cS =  \frac{(\Delta t)^2}{2} \!\sum_{m,n=1}^N \! \left(f(x_{m-1}) - v_m \right) G_{m,n}(g) \left(f(x_{n-1}) - v_n \right),
	\end{equation}
	omitting $f$-independent terms. 
	This bilinear action corresponds to a Gaussian field theory \cite{Zinn-Justin2002} in $f(x)$ of mean $\avg{f(x_{n-1})} = v(x_n)$ with non-local correlated fluctuations following $\avg{f(x_{m-1}) f(x_{n-1})} - v_m v_n = (G^{-1})_{m,n}$. 
	
	Parametrising $f$ polynomially, $f(x) = \sum_{\ell=0}^{L-1} f_{\ell} x^{\ell}$ for some $L < N$, the optimal coefficients $\hat{f}_{\ell}$ can be found by inserting the polynomial ansatz into \Eqref{eq:log_likelihood_f} and setting $\partial \cS / \partial \hat{f}_{\ell} = 0$. The resulting action in the $f_{\ell}$ resembles the Hamiltonian of $L$ fully coupled harmonic oscillators in a confining harmonic potential which allows for a single optimal configuration corresponding to the estimate $\hat{f}_{\ell}$; a calculation provided in \ref{append:fixed_diffusion} shows that this minimum is given by $\hat{f}_{\ell} = (H^{-1} \vec{j})_{\ell}$, where we introduced 
	\begin{equation}
	H_{\ell,k} = \sum_{m,n}x_{m-1}^{\ell}G_{m,n}x_{n-1}^{k} \qquad \mathrm{and} \qquad j_{\ell}=\sum_{m,n} G_{m,n} v_n x_{m-1}^{\ell} \ .
	\end{equation} 
	This estimate is exact for arbitrary inhomogeneous fixed diffusion $g(x)$; it further simplifies when $f$ is linear and $g$ constant (discretised fractional Ornstein-Uhlenbeck process, see \ref{append:FOU}).
	
	Analogously, we consider the case of $f(x)$ general but fixed. Introducing the \emph{empirical noise correlation} 
	\begin{equation}
	[D(f)]_{m,n} = \left(f(x_{m-1}) - v_{m}\right) C_{m,n}^{-1} \left(f(x_{n-1}) - v_{n}\right)
	\end{equation}
	together with the inverse field $\psi_{n} = (g(x_{n-1}))^{-1}$, $\cS$  reads
	\begin{equation}
	\label{eq:log_likelihood_g}
	\cS = \frac{(\Delta t)^2}{2} \sum_{m,n=1}^N \psi_m D_{m,n}(f) \psi_n - \sum_{n=1}^N \ln \left| \psi_n \right| . 
	\end{equation}
	The action resembles a fully interacting $N$-body Hamiltonian on the positive half line; 
	the ``particles'' at position $\psi_n$ are quadratically coupled to one another, confined harmonically for $\psi_n \to \infty$, since $D_{nn} > 0$, yet repelled from the origin by a logarithmically diverging potential giving rise to a stable minimum satisfying a self-consistency relation 
	\begin{equation}
	(\Delta t)^2\hat{\psi}_n\sum_{m}D_{m,n}\hat{\psi}_{m} = 1
	\end{equation} for all $n$.
	If $g \equiv g_0$ is assumed to be constant (additive noise), this immediately returns $\hat{g}^2_0 = (\Delta t)^2 \sum_{m,n}D_{m,n}$. 
	If further $H=\frac{1}{2}$, $D_{m,n}$ is diagonal, and \Eqref{eq:log_likelihood_g} resembles the Hamiltonian of $N$ non-interacting particles in a logarithmic-harmonic potential \cite{Ryabov2013}. The estimate for $g_0$ then recovers the well-known Markovian result \cite{Kleinhans2012}
	\begin{equation}
	\hat{g}^2_0 = \sum_{n} \left( f(x_{n-1})\Delta t - (x_n- x_{n-1})\right)^2 \,.
	\end{equation}

	\begin{figure*}[t]
		\centering
		\includegraphics[width=\textwidth]{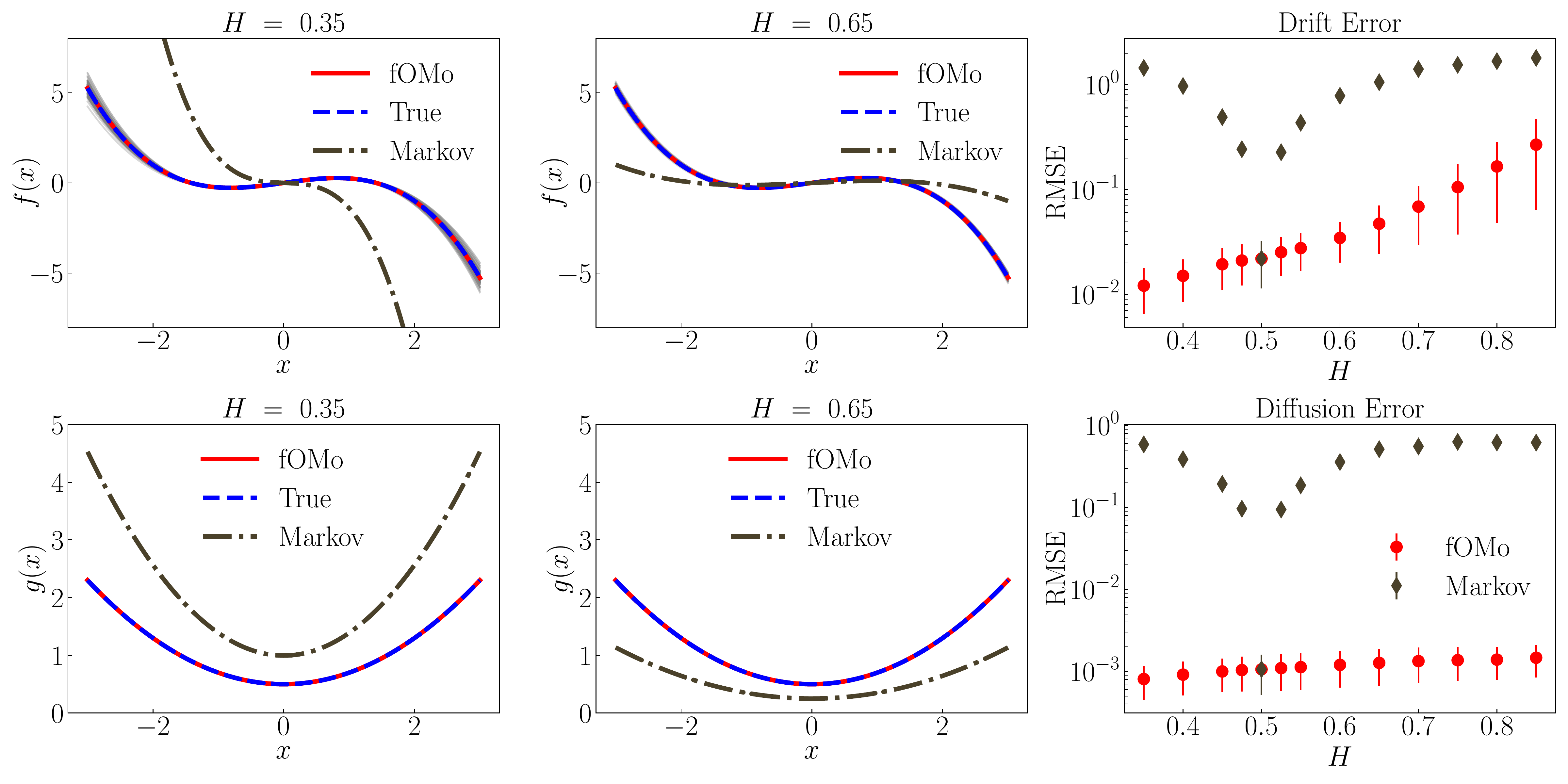}
		\caption{Inference via fOMo correctly recovers drift and diffusion in contrast to Markovian estimation. Ensemble study with $100$ trajectories with $N=10^6, \Delta t = 10^{-2}$, random initial conditions and equilibration time $T_{\mathrm{eq}} = 10^4$ \cite{Davies1987}. (top left and top center panels) Drift and (bottom left and bottom center panels) diffusion estimation via fOMo (red) and Markov least-squares fit (brown) for anti-correlated ($H=0.35$) and long-range correlated ($H=0.65$) dynamics. Inferred drift and diffusion of all ensemble members are plotted in light grey, only visible in (top left panel). (top right panel) Reconstruction error of drift $f(x)$ for fOMo (red) and Markovian estimate (brown).  (bottom right panel) Reconstruction error of diffusion $g(x)$. Root-mean-square-errors are computed using the empirical invariant density (see text).}
		\label{fig:synthetic}
	\end{figure*}

	\section{Superposition of noise processes} 
	The method is readily adapted to processes subject to different additional Gaussian noise sources which could model, for instance, the additional coupling of the process to a heat bath. 
	We consider a generalisation of $x_n$, the fractional process introduced in \Eqref{eq:FLE_def}, by setting 
	\begin{equation}
	{x_{n+1} = x_n  + f(x_n) \dt + g(x_n) \dWh_{n+1} + \eta_{n+1}}\ ,
	\end{equation} 
	where $\eta_n$ is an additional independent Gaussian noise term of correlation $K_{m,n} = \overline{\eta_m \eta_n}$. 
	Repeating the previous steps in constructing the fractional Onsager-Machlup action, one inverts the new stochastic equation for $x_n$ to find 
	\begin{equation}
	\left[\vec{\dWh}(\x)\right]_n = (x_n - x_{n-1} - f(x_{n-1})\Delta t - \eta_n)/g(x_n)\ ,
	\end{equation}
	which one inserts into the free action $\cA[\vec{\dWh}]$ (see Equations~(\ref{eq:emp_noise}, \ref{eq:log_likelihood})). This transformation leaves the Jacobian, $\sum_n \left|\left( g(x_n) \right)\right|^{-1}$, unchanged. Instead the path  likelihood of $\x$ conditioned on a particular $\vec{\eta}$, \Eqref{eq:log_likelihood}, acquires some extra terms of the form  
	$\sum_n \eta_n a_n - \frac{1}{2} \sum_{m,n} \eta_m G_{m,n}\eta_n$, where we abbreviate $a_n = \sum_m G_{m,n} \left(f(x_{m-1}) - v_m \right)\Delta t$. It remains to average over the additional noise distribution 
	\begin{equation}
	P_{\eta}[\vec{\eta}] = \left( (2 \pi)^N \det K \right)^{-\frac{1}{2}}\exp\left( - 1/2 \sum_{m,n} \eta_m (K^{-1})_{mn} \eta_n \right) \ .
	\end{equation}
	Carrying out the Gaussian integral in $\prod_n \dint{\eta_n}$, one finds that the modified action is
	\begin{eqnarray}
	\fl \widetilde{\cS} &= -\ln\overline{ P\left[\x|f,g\right]}=-\ln \int D[\vec{\eta}]P[\x|f,g,\vec{\eta}]P_{\eta}[\vec{\eta}] \\
	\fl    &= \frac{(\Delta t)^2}{2} \sum_{m,n=1}^N \left(f(x_{m-1})- v_m \right) \widetilde{G}_{m,n}\left(f(x_{m-1})- v_m \right) 
	+ \sum_{n=0}^{N-1} \ln \left| g_n\right|,
	\label{eq:disorder_avg_fOM}
	\end{eqnarray}
	where we omit terms independent of $f,g$, and, in contrast to \Eqref{eq:log_likelihood},  replace the empirical propagator by ${\widetilde{G} = [ G^{-1} + K ]^{-1}}$.

	\section{Fast Inversion}
	In order to numerically evaluate the log-likelihood given by \Eqref{eq:log_likelihood} the correlation matrix $C$ has to be inverted. This is an operation of computational complexity $\mathcal{O}(N^3)$ and memory requirements of order $\mathcal{O}(N^2)$, 
	rendering it prohibitively costly for times series of only intermediate length ($N\sim 10^4$). We circumvent this problem by exploiting the Toeplitz structure of the correlation matrix;
	since the process is stationary,
	$C_{mn}$ depends on $|m-n|$ only  and can be inverted efficiently \cite{Lv2007}.
	In doing so, we reduce computational complexity to $\cO(N^2)$ and memory requirements to $\cO(N)$ (see \ref{appendix:numerics} for further details), 
	rendering fOMo computationally tractable even for long time series ($N \sim 10^6$). 
	This fast inversion method may be generalised to other stationary driving processes such as, for instance,
	tempered fractional noise \cite{Meerschaert2013, Liemert2017, Molina-Garcia2018}.

	\section{Synthetic Data}
	We have successfully tested fOMo with various  nonlinear models and illustrate a representative example in the following.
	We consider a model system defined by \Eqref{eq:FLE_def} choosing ${f(x) = -0.25 x^3 + 0.5 x}$, corresponding to a double well potential, and inhomogeneous diffusion ${g(x) = 0.2 x^2 + 0.5}$, implying large stochastic fluctuations far away from the origin. We assume the Hurst exponent to be already determined and infer $f$ and $g$ employing fOMo (\Eqref{eq:log_likelihood}).
	In order to highlight the significance of the noise correlations, we compare this estimate to a fOMo estimate where we \emph{wrongly} fix $H=\frac{1}{2}$, whence $C^{-1}$ is diagonal, effectively producing a Markovian maximum likelihood estimate (see \cite{Kleinhans2012}).
	We use polynomial basis functions for drift ($L=3$) and diffusion ($L=2$). For an ensemble of $100$ trajectories (see Figure~\ref{fig:synthetic} for details), we simultaneously infer drift and diffusion terms and measure the root-mean-square error (RMSE) of the inferred terms $\hat{f}, \hat{g}$. Inference errors are weighted with the empirical invariant measure of the process 
	\begin{equation}
	\mathrm{RMSE}(\hat{f}) = \left[\sum_{n=1}^N \left(\hat{f}(x_n) - f(x_n)\right)^2/N\right]^{\frac{1}{2}}, 
	\end{equation}
	and analogous for $g$.

	Remarkably, the Markovian model underestimates (overestimates) both drift and diffusion terms for $H>\frac{1}{2}$ ($H<\frac{1}{2}$) while fOMo correctly recovers the true input functions (Figure~\ref{fig:synthetic}) with excellent accuracy. The (anti-)correlations of the noise process lead to a wider (narrower) width of the invariant density of the process, also explaining the visible deviations of the fOMo samples from the ensemble mean for $H=0.35$ (Figure~\ref{fig:synthetic} (top left panel)). Markovian modelling fails to recover the double well shape of the corresponding potential in our example for $H=0.35$. The inference error of Markovian modelling steeply increases for small deviations from $H=\frac{1}{2}$, highlighting the necessity of taking fractional correlations into account.
	
	We investigate the finite-size error scaling as a function of the trajectory length $N$ and find a drift error scaling $\mathrm{RMSE}\sim 1/\sqrt{N}$ for $H=\frac{1}{2}$, and $\mathrm{RMSE}\sim N^{\alpha}$ with $\alpha \sim H^2$ for $H\neq\frac{1}{2}$, while the diffusion error scaling does not show a clear trend (see \ref{append:error_scaling}).

	\section{Temperature Data}
	We apply fOMo to daily mean temperature data recorded at Potsdam Telegrafenberg weather station, Germany. This time series consists of 130 years of uninterrupted temperature measurements downloaded from the ECAD project \cite{ECAD, Besselaar2015}. Removing the seasonal cycle using a Fourier series, we obtain the temperature anomalies $T$, an approximately stationary time series. Temperature anomalies are long-range correlated \cite{Bunde2001, Eichner2003}, monofractal \cite{Bunde2017} and have been described by overdamped models driven by fractional Gaussian noise \cite{Massah2016, Janosi2002, Kassel2022}. We determine the Hurst exponent using detrended fluctuation analysis with a cubic polynomial (DFA3) \cite{Peng1994, Hoell2019} and construct the correlation matrix using the estimated Hurst exponent $H \approx 0.65$ \cite{Massah2016}, and a sampling time $\Delta t = 1\,\mathrm{d}$.
	\begin{figure}
		\centering
		\includegraphics[width=0.495\textwidth]{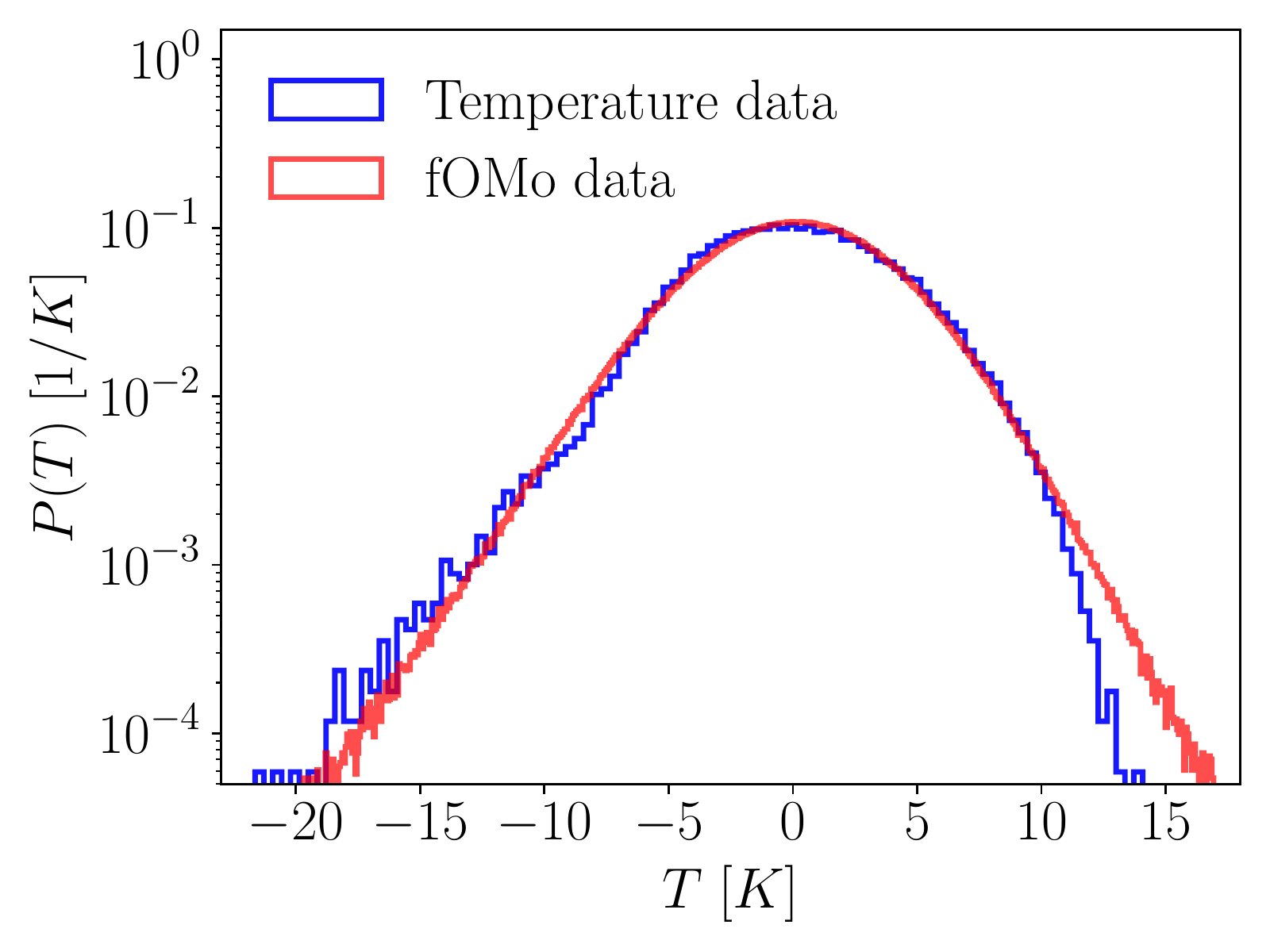}
		\includegraphics[width=0.49\textwidth]{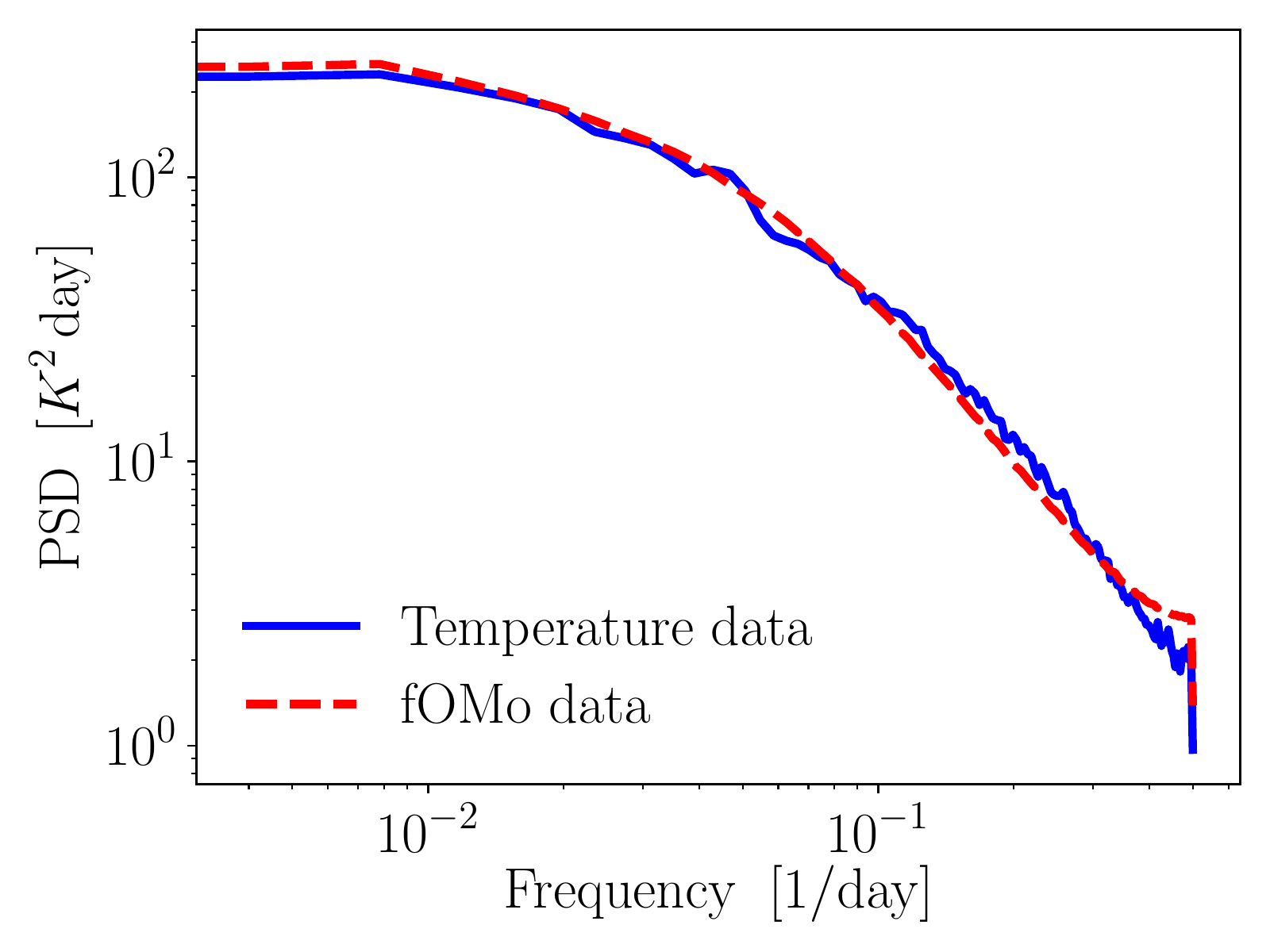}
		\includegraphics[width=0.49\textwidth]{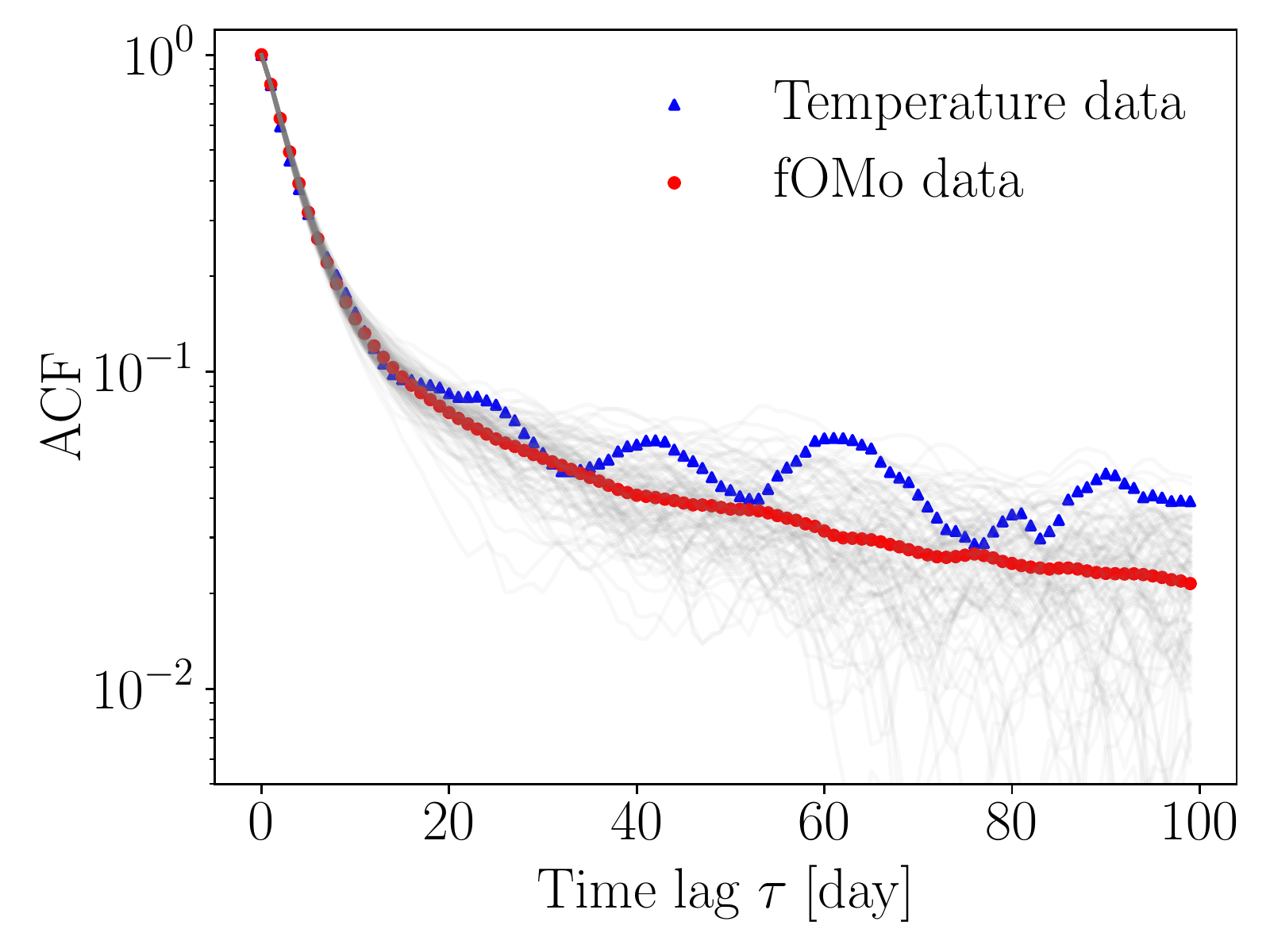}
		\includegraphics[width=0.49\textwidth]{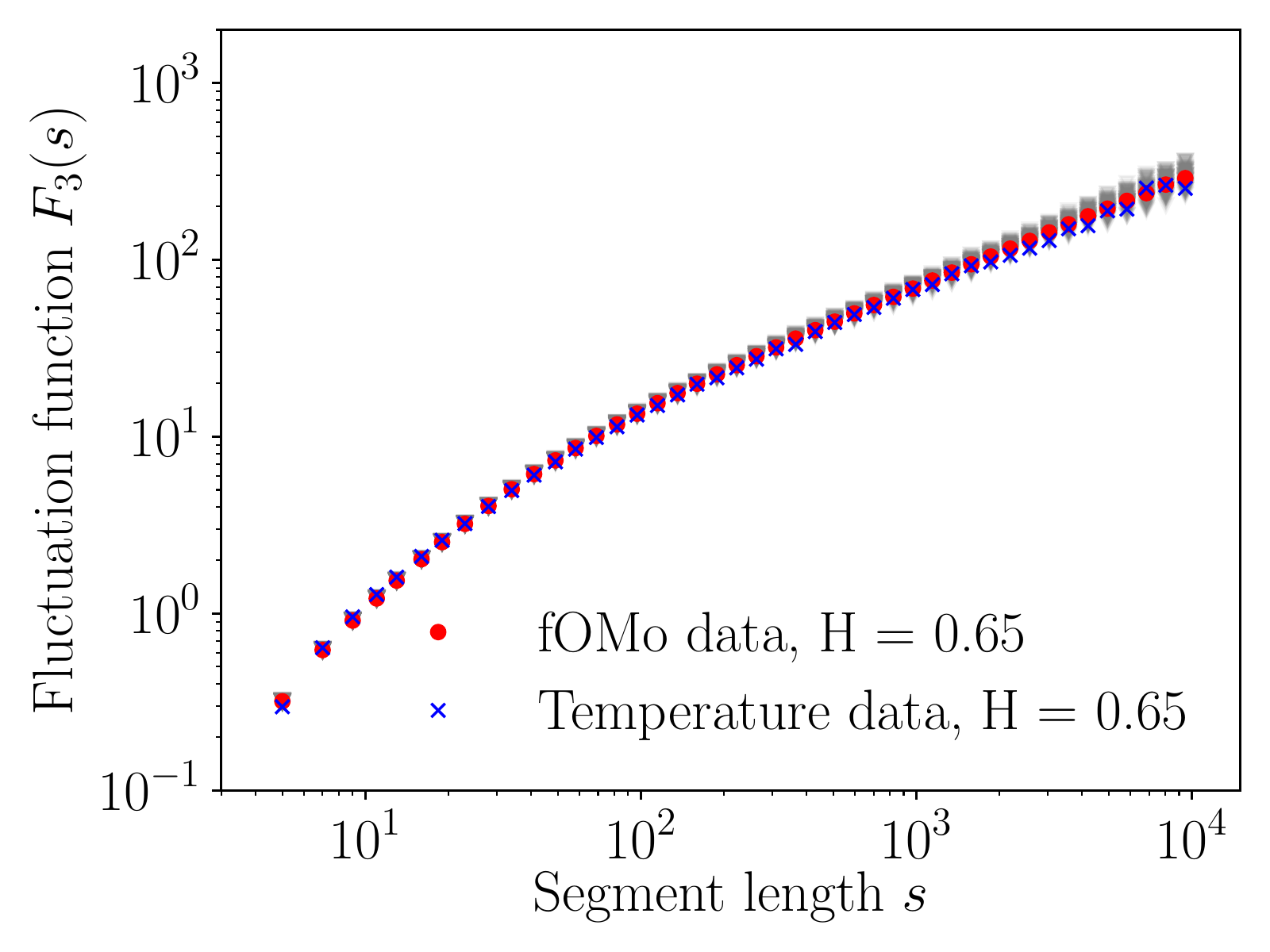}
		\caption{Stochastic fractional nonlinear model inferred with fOMo for Potsdam daily mean temperature anomalies compared to original data. Semi-transparent grey lines indicate statistics of $100$ model data trajectories, red lines are ensemble averages of model data (labelled fOMo data) and blue lines show statistics of original data. Top left: Histogram of marginal distribution. Top right: Power spectral density (PSD). Bottom left: Autocorrelation function (ACF). Bottom right: DFA3 fluctuation function.}
		\label{fig:potsdam_comparison}
	\end{figure}
	\begin{figure}[h]
		\includegraphics[width=1.0\textwidth]{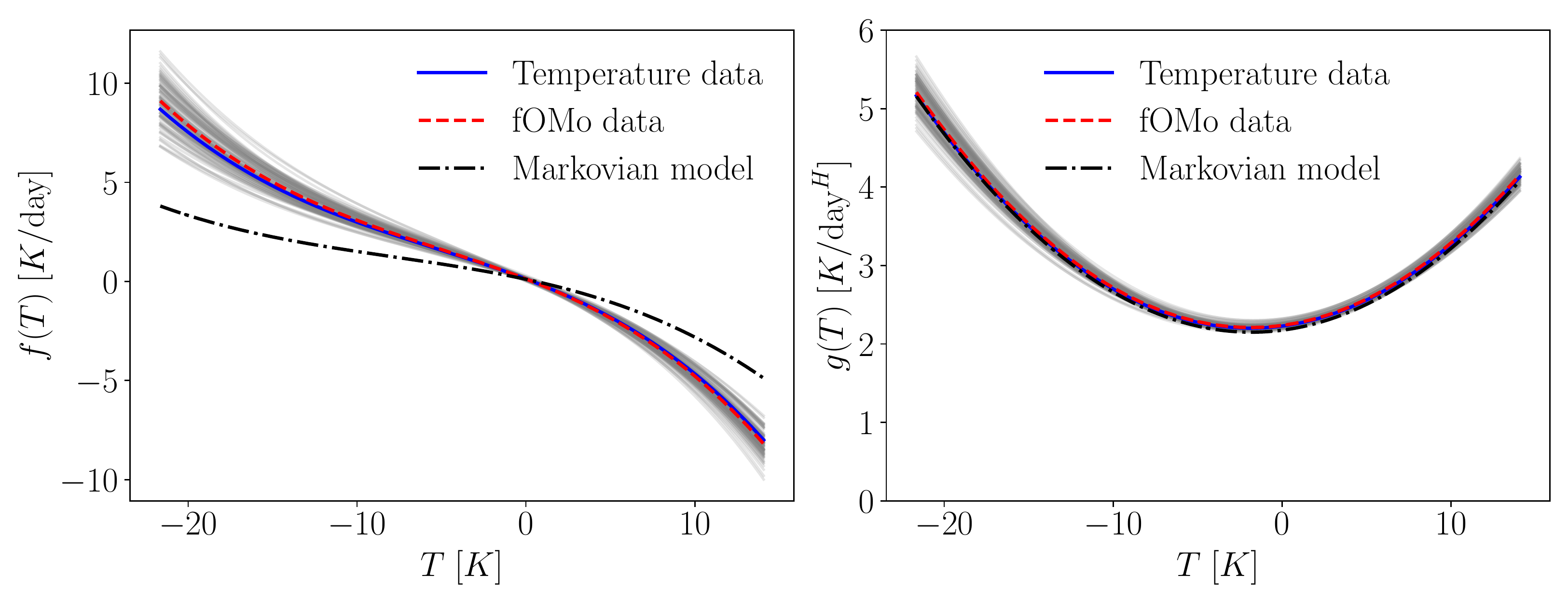}
		\caption{Drift and diffusion term estimates for Potsdam daily mean temperature anomalies. Blue lines indicate fOMo estimates from the original data and dark olive lines the Markovian estimate from original data. Semi-transparent lines show $100$ fOMo estimates from model data generated using \Eqref{eq:FLE_def} and inferred drift and diffusion (blue lines). Red lines indicate the ensemble mean of the grey semi-transparent lines. Red and blue lines show excellent agreement, indicating that the estimator is free from bias.}
		\label{fig:potsdam_drift_diff}
	\end{figure}
	In this model, $f$ may be interpreted as an atmospheric response function bearing the unit $[f] = K/\mathrm{day}$ and its stochastic fluctuations $g$, measured in $[g]=K/\mathrm{day}^H$ \cite{Janosi2002}. We use cubic and quadratic polynomial ansatzes for drift and diffusion, respectively and employ fOMo and a Markovian maximum likelihood estimate ($H=\frac{1}{2}$, see above). Since $f$ and $g$ of real-world data are not \emph{a priori} known, we cannot conduct an error analysis as described above for synthetic data. However, we can compare model data with the original data and check the consistency of the model. To this end, we generate an ensemble of model time series using \Eqref{eq:FLE_def} and the parameters of $f$ and $g$ obtained via fOMo. Subsequently, we compare statistics of the original data and the model data ensemble. Additionally, we infer drift and diffusion parameters from the synthetic model data ensemble and compare these with drift and diffusion obtained for the original data. For a method without bias, the mean of the inferred drift and diffusion terms should coincide with the drift and diffusion estimates of the original data. The spread of the ensemble of inferred drift and diffusion terms then gives an estimate of the errors of the inferred drift and diffusion terms due to finite samples and thus hints at the reliability of the estimate. Temperature data and synthetic model data are in very good agreement, as Figure~\ref{fig:potsdam_comparison} shows. Furthermore, the model is consistent since the mean of inferred drift and diffusion from model data coincides with drift and diffusion estimates from the original data (see Figure~\ref{fig:potsdam_drift_diff}). The Markovian model significantly underestimates the deterministic force term compared to the force term inferred by fOMo which takes the long-range correlations into account. Unlike for the double well potential, the Markovian diffusion estimate agrees well with the fOMo estimate. This is due to the approximate linearity of $f$.
	In this article, we neglect the estimation error of the Hurst exponent. However, we propose the following procedure using a recently published operational method for identifying scaling regimes in attractor dimension estimation \cite{Desmukh2021} which may also be employed to obtain error bars for Hurst exponents estimated via DFA \cite{Phillips2023}. At first, one determines the Hurst exponent values at the lower and upper error bars and subsequently conducts fOMo with these values. The obtained parameters for drift and diffusion terms then serve as an error estimate.
	
	\section{Conclusion}
	By way of a fractional generalisation of the Onsager-Machlup formalism, we create an estimation method that is able to infer functional parameters of a fully nonlinear model driven by multiplicative fractional noise from a single trajectory. Applying the algorithm to both synthetic and real data, we recover excellent estimates even for Hurst values deep in the non-Markovian regime, where ignoring the (anti-)correlations of the fluctuations leads to gross systematic errors. The method provides a needed tool for modelling real-world complex systems whose fractal nature cannot be neglected, and illuminates intriguing connections bridging time series analysis with the statistical physics of fractionally correlated many-body systems.
	
	\ack
	JAK thanks SISSA for hospitality, Stefano Bo for useful discussions, and Steffen Peters for IT support.	
	
	BW thanks the MPIPKS for hospitality.
	
	BW acknowledges financial support from the Imperial College Borland Research Fellowship and from the MIUR PRIN project ``Coarse-grained description for non-equilibrium systems and transport phenomena (CO-NEST)'' n. 201798CZL. BW further acknowledges support from SISSA, where a part of this work was carried out, and INFN.
	
	

\begin{thebibliography}{10}
		\expandafter\ifx\csname url\endcsname\relax
		\def\url#1{{\tt #1}}\fi
		\expandafter\ifx\csname urlprefix\endcsname\relax\def\urlprefix{URL }\fi
		\providecommand{\eprint}[2][]{\url{#2}}
		
		\bibitem{Metzler2014}
		Metzler R, Jeon J~H, Cherstvy A~G and Barkai E 2014 {\em Phys. Chem. Chem.
			Phys.\/} {\bf 16}(44) 24128--24164
		\urlprefix\url{http://dx.doi.org/10.1039/C4CP03465A}
		
		\bibitem{Hurst1951}
		Hurst H~E 1951 {\em Trans. Am. Soc. Civ. Eng.\/} {\bf 116} 770
		
		\bibitem{Bunde2001}
		Bunde A, Havlin S, Koscielny-Bunde E and Schellnhuber H~J 2001 {\em Physica A:
			Statistical Mechanics and its Applications\/} {\bf 302} 255--267 ISSN
		0378-4371
		\urlprefix\url{https://www.sciencedirect.com/science/article/pii/S0378437101004691}
		
		\bibitem{Eichner2003}
		Eichner J~F, Koscielny-Bunde E, Bunde A, Havlin S and Schellnhuber H~J 2003
		{\em Phys. Rev. E\/} {\bf 68}(4) 046133
		\urlprefix\url{https://link.aps.org/doi/10.1103/PhysRevE.68.046133}
		
		\bibitem{Bunde2005}
		Bunde A, Eichner J~F, Kantelhardt J~W and Havlin S 2005 {\em Phys. Rev.
			Lett.\/} {\bf 94}(4) 048701
		\urlprefix\url{https://link.aps.org/doi/10.1103/PhysRevLett.94.048701}
		
		\bibitem{Kantelhardt2006}
		Kantelhardt J~W, Koscielny-Bunde E, Rybski D, Braun P, Bunde A and Havlin S
		2006 {\em J. Geophys. Res. Atmos.\/} {\bf 111}
		\urlprefix\url{https://agupubs.onlinelibrary.wiley.com/doi/abs/10.1029/2005JD005881}
		
		\bibitem{Voss1992}
		Voss R~F 1992 {\em Phys. Rev. Lett.\/} {\bf 68} 3805
		
		\bibitem{Chatzidimitriou1993}
		Chatzidimitriou-Dreismann C and Larhammar D 1993 {\em Nature\/} {\bf 361}
		212--213
		
		\bibitem{Hofling2013}
		Höfling F and Franosch T 2013 {\em Rep. Prog. Phys.\/} {\bf 76} 046602 ISSN
		0034-4885 \urlprefix\url{https://dx.doi.org/10.1088/0034-4885/76/4/046602}
		
		\bibitem{Wan2014}
		Wan K~Y and Goldstein R~E 2014 {\em Phys. Rev. Lett.\/} {\bf 113}(23) 238103
		\urlprefix\url{https://link.aps.org/doi/10.1103/PhysRevLett.113.238103}
		
		\bibitem{Palva2013}
		Palva J~M, Zhigalov A, Hirvonen J, Korhonen O, Linkenkaer-Hansen K and Palva S
		2013 {\em Proc. Natl. Acad. Sci. U.S.A.\/} {\bf 110} 3585--3590
		
		\bibitem{Teka2014}
		Teka W, Marinov T~M and Santamaria F 2014 {\em {PLoS} Comput. Biol.\/} {\bf 10}
		e1003526 ISSN 1553-7358
		\urlprefix\url{https://dx.plos.org/10.1371/journal.pcbi.1003526}
		
		\bibitem{Munoz-Gil2021}
		Mu\~{n}oz Gil G, Volpe G, Garcia-March M~A, Aghion E, Argun A, Hong C~B, Bland
		T, Bo S, Conejero J~A, Firbas N, Garibo~i Orts O, Gentili A, Huang Z, Jeon
		J~H, Kabbech H, Kim Y, Kowalek P, Krapf D, Loch-Olszewska H, Lomholt M~A,
		Masson J~B, Meyer P~G, Park S, Requena B, Smal I, Song T, Szwabi\'{n}ski J,
		Thapa S, Verdier H, Volpe G, Widera A, Lewenstein M, Metzler R and Manzo C
		2021 {\em Nat. Commun.\/} {\bf 12} 6253 ISSN 2041-1723
		\urlprefix\url{https://doi.org/10.1038/s41467-021-26320-w}
		
		\bibitem{Bzdok2018}
		Bzdok D, Altman N and Krzywinski M 2018 {\em Nat. Methods\/} {\bf 15} 233--234
		ISSN 1548-7105 \urlprefix\url{https://doi.org/10.1038/nmeth.4642}
		
		\bibitem{Carleo2019}
		Carleo G, Cirac I, Cranmer K, Daudet L, Schuld M, Tishby N, Vogt-Maranto L and
		Zdeborov\'a L 2019 {\em Rev. Mod. Phys.\/} {\bf 91}(4) 045002
		\urlprefix\url{https://doi.org/10.1103/RevModPhys.91.045002}
		
		\bibitem{Cichos2020}
		Cichos F, Gustavsson K, Mehlig B and Volpe G 2020 {\em Nature Machine
			Intelligence\/} {\bf 2} 94--103 ISSN 2522-5839
		\urlprefix\url{https://doi.org/10.1038/s42256-020-0146-9}
		
		\bibitem{Rao2011}
		Rao B~P 2011 {\em Statistical inference for fractional diffusion processes\/}
		(Chichester, UK: John Wiley \& Sons)
		\urlprefix\url{https://www.wiley.com/en-gb/Statistical+Inference+for+Fractional+Diffusion+Processes-p-9780470975763}
		
		\bibitem{Burnecki2014}
		Burnecki K and Weron A 2014 {\em J. Stat. Mech.: Theory Exp.\/} {\bf 2014}
		P10036
		
		\bibitem{Graves2015}
		Graves T, Gramacy R~B, Franzke C~L and Watkins N~W 2015 {\em Nonlinear Process
			Geophys\/} {\bf 22} 679--700
		
		\bibitem{Kubilius2017}
		Kubilius K, Mishura Y and Ralchenko K 2017 {\em Parameter Estimation in
			Fractional Diffusion Models\/} (Cham, Switzerland: Springer)
		\urlprefix\url{https://doi.org/10.1007/978-3-319-71030-3}
		
		\bibitem{Thapa2018}
		Thapa S, Lomholt M~A, Krog J, Cherstvy A~G and Metzler R 2018 {\em Phys. Chem.
			Chem. Phys.\/} {\bf 20}(46) 29018--29037
		\urlprefix\url{http://dx.doi.org/10.1039/C8CP04043E}
		
		\bibitem{Siegert1998}
		Siegert S, Friedrich R and Peinke J 1998 {\em Phys. Lett. A\/} {\bf 243} 275 --
		280 ISSN 0375-9601
		\urlprefix\url{http://www.sciencedirect.com/science/article/pii/S0375960198002837}
		
		\bibitem{Ragwitz2001}
		Ragwitz M and Kantz H 2001 {\em Phys. Rev. Lett.\/} {\bf 87}(25) 254501
		\urlprefix\url{https://link.aps.org/doi/10.1103/PhysRevLett.87.254501}
		
		\bibitem{Boettcher2006}
		B\"ottcher F, Peinke J, Kleinhans D, Friedrich R, Lind P~G and Haase M 2006
		{\em Phys. Rev. Lett.\/} {\bf 97}(9) 090603
		\urlprefix\url{https://link.aps.org/doi/10.1103/PhysRevLett.97.090603}
		
		\bibitem{Beheiry2015}
		Beheiry M~E, Dahan M and Masson J~B 2015 {\em Nat. Methods\/} {\bf 12} 594--595
		ISSN 1548-7105 \urlprefix\url{https://doi.org/10.1038/nmeth.3441}
		
		\bibitem{Gouasmi2017}
		Gouasmi A, Parish E~J and Duraisamy K 2017 {\em Proceedings of the Royal
			Society A: Mathematical, Physical and Engineering Sciences\/} {\bf 473}
		20170385 (\textit{Preprint}
		\eprint{https://royalsocietypublishing.org/doi/pdf/10.1098/rspa.2017.0385})
		\urlprefix\url{https://royalsocietypublishing.org/doi/abs/10.1098/rspa.2017.0385}
		
		\bibitem{Lehle2018}
		Lehle B and Peinke J 2018 {\em Phys. Rev. E\/} {\bf 97}(1) 012113
		\urlprefix\url{https://link.aps.org/doi/10.1103/PhysRevE.97.012113}
		
		\bibitem{PerezGarcia2018}
		P\'erez~Garc\'ia L, Donlucas~P\'erez J, Volpe G, V~Arzola A and Volpe G 2018
		{\em Nat. Commun.\/} {\bf 9} 5166 ISSN 2041-1723
		\urlprefix\url{https://doi.org/10.1038/s41467-018-07437-x}
		
		\bibitem{Tabar2019Book}
		Tabar R 2019 {\em Analysis and data-based reconstruction of complex nonlinear
			dynamical systems\/} vol 730 (Cham, Switzerland: Springer)
		\urlprefix\url{https://link.springer.com/book/10.1007/978-3-030-18472-8}
		
		\bibitem{Brueckner2020}
		Br\"uckner D~B, Ronceray P and Broedersz C~P 2020 {\em Phys. Rev. Lett.\/} {\bf
			125}(5) 058103
		\urlprefix\url{https://link.aps.org/doi/10.1103/PhysRevLett.125.058103}
		
		\bibitem{Frishman2020}
		Frishman A and Ronceray P 2020 {\em Phys. Rev. X\/} {\bf 10}(2) 021009
		\urlprefix\url{https://link.aps.org/doi/10.1103/PhysRevX.10.021009}
		
		\bibitem{Ferretti2020}
		Ferretti F, Chard\`es V, Mora T, Walczak A~M and Giardina I 2020 {\em Phys.
			Rev. X\/} {\bf 10}(3) 031018
		\urlprefix\url{https://link.aps.org/doi/10.1103/PhysRevX.10.031018}
		
		\bibitem{Willers2021}
		Willers C and Kamps O 2021 {\em Eur. Phys. J. B\/} {\bf 94} 149 ISSN 1434-6036
		\urlprefix\url{https://doi.org/10.1140/epjb/s10051-021-00149-0}
		
		\bibitem{Weiss2013}
		Weiss M 2013 {\em Phys. Rev. E\/} {\bf 88} 010101
		\urlprefix\url{https://link.aps.org/doi/10.1103/PhysRevE.88.010101}
		
		\bibitem{Saelens2019}
		Saelens W, Cannoodt R, Todorov H and Saeys Y 2019 {\em Nat Biotechnol\/} {\bf
			37} 547--554 ISSN 1546-1696
		\urlprefix\url{https://www.nature.com/articles/s41587-019-0071-9}
		
		\bibitem{Onsager1953}
		Onsager L and Machlup S 1953 {\em Phys. Rev.\/} {\bf 91}(6) 1505--1512
		\urlprefix\url{https://link.aps.org/doi/10.1103/PhysRev.91.1505}
		
		\bibitem{Durr1978}
		Dürr D and Bach A 1978 {\em Commun. Math. Phys.\/} {\bf 60} 153--170 ISSN
		1432-0916 \urlprefix\url{https://doi.org/10.1007/BF01609446}
		
		\bibitem{Taeuber2014}
		T{\"a}uber U~C 2014 {\em Critical dynamics\/} (Cambridge, UK: Cambridge
		University Press)
		
		\bibitem{Cugliandolo2017}
		Cugliandolo L~F and Lecomte V 2017 {\em J. Phys. A Math. Theor.\/} {\bf 50}
		345001 \urlprefix\url{https://dx.doi.org/10.1088/1751-8121/aa7dd6}
		
		\bibitem{Pipiras2017}
		Pipiras V and Taqqu M~S 2017 {\em Long-Range Dependence and Self-Similarity\/}
		Cambridge Series in Statistical and Probabilistic Mathematics (Cambridge
		University Press)
		
		\bibitem{Mandelbrot1968}
		Mandelbrot B~B and Van~Ness J~W 1968 {\em SIAM Rev.\/} {\bf 10} 422--437
		
		\bibitem{Mandelbrot1969}
		Mandelbrot B~B and Wallis J~R 1969 {\em Water Resour. Res\/} {\bf 5} 321--340
		
		\bibitem{Peng1994}
		Peng C~K, Buldyrev S~V, Havlin S, Simons M, Stanley H~E and Goldberger A~L 1994
		{\em Phys. Rev. E\/} {\bf 49}(2) 1685--1689
		\urlprefix\url{https://link.aps.org/doi/10.1103/PhysRevE.49.1685}
		
		\bibitem{Hoell2019}
		H\"oll M, Kiyono K and Kantz H 2019 {\em Phys. Rev. E\/} {\bf 99}(3) 033305
		\urlprefix\url{https://link.aps.org/doi/10.1103/PhysRevE.99.033305}
		
		\bibitem{Abry1998}
		Abry P and Veitch D 1998 {\em {IEEE} Trans. Inf. Theory\/} {\bf 44} 2--15
		
		\bibitem{Bo2019}
		Bo S, Schmidt F, Eichhorn R and Volpe G 2019 {\em Phys. Rev. E\/} {\bf 100}(1)
		010102 \urlprefix\url{https://link.aps.org/doi/10.1103/PhysRevE.100.010102}
		
		\bibitem{Devore2012}
		Devore J~L, Berk K~N and Carlton M~A 2021 {\em Modern Mathematical Statistics
			with Applications\/} 3rd ed Springer Texts in Statistics (Cham, Switzerland:
		Springer) ISBN 9781461403906,1461403901
		
		\bibitem{Zinn-Justin2002}
		Zinn-Justin J 2002 {\em Quantum Field Theory and Critical Phenomena\/} (Oxford
		University Press) ISBN 978-0-19-850923-3
		\urlprefix\url{https://academic.oup.com/book/11638}
		
		\bibitem{Horsthemke1975}
		Horsthemke W and Bach A 1975 {\em Z Physik B\/} {\bf 22} 189--192
		
		\bibitem{Adib2008}
		Adib A~B 2008 {\em J. Phys. Chem. B\/} {\bf 112} 5910--5916
		
		\bibitem{Kappler2020}
		Kappler J and Adhikari R 2020 {\em Phys. Rev. Res.\/} {\bf 2}(2) 023407
		\urlprefix\url{https://link.aps.org/doi/10.1103/PhysRevResearch.2.023407}
		
		\bibitem{Gladrow2021}
		Gladrow J, Keyser U~F, Adhikari R and Kappler J 2021 {\em Phys. Rev. X\/} {\bf
			11}(3) 031022
		\urlprefix\url{https://link.aps.org/doi/10.1103/PhysRevX.11.031022}
		
		\bibitem{Kieninger2021}
		Kieninger S and Keller B~G 2021 {\em J. Chem. Phys.\/} {\bf 154} 094102
		
		\bibitem{Kleinhans2012}
		Kleinhans D 2012 {\em Phys. Rev. E\/} {\bf 85}(2) 026705
		\urlprefix\url{https://link.aps.org/doi/10.1103/PhysRevE.85.026705}
		
		\bibitem{Ryabov2013}
		Ryabov A, Dierl M, Chvosta P, Einax M and Maass P 2013 {\em J. Phys. A: Math.
			Theor.\/} {\bf 46} 075002 ISSN 1751-8121
		
		\bibitem{Davies1987}
		Davies R~B and Harte D~S 1987 {\em Biometrika\/} {\bf 74} 95--101 ISSN
		0006-3444 \urlprefix\url{https://doi.org/10.1093/biomet/74.1.95}
		
		\bibitem{Lv2007}
		Lv X~G and Huang T~Z 2007 {\em Appl. Math. Lett.\/} {\bf 20} 1189--1193 ISSN
		0893-9659
		\urlprefix\url{https://www.sciencedirect.com/science/article/pii/S0893965907000535}
		
		\bibitem{Meerschaert2013}
		Meerschaert M~M and Sabzikar F 2013 {\em Stat. Probab. Lett\/} {\bf 83}
		2269--2275 ISSN 0167-7152
		
		\bibitem{Liemert2017}
		Liemert A, Sandev T and Kantz H 2017 {\em Phys. A: Stat. Mech. Appl.\/} {\bf
			466} 356--369
		
		\bibitem{Molina-Garcia2018}
		Molina-Garcia D, Sandev T, Safdari H, Pagnini G, Chechkin A and Metzler R 2018
		{\em New J. Phys.\/} {\bf 20} 103027
		\urlprefix\url{https://iopscience.iop.org/article/10.1088/1367-2630/aae4b2/meta}
		
		\bibitem{ECAD}
		Klein~Tank A~M~G, Wijngaard J~B, Können G~P, Böhm R, Demarée G, Gocheva A,
		Mileta M, Pashiardis S, Hejkrlik L, Kern-Hansen C, Heino R, Bessemoulin P,
		Müller-Westermeier G, Tzanakou M, Szalai S, Pálsdóttir T, Fitzgerald D,
		Rubin S, Capaldo M, Maugeri M, Leitass A, Bukantis A, Aberfeld R, van Engelen
		A~F~V, Forland E, Mietus M, Coelho F, Mares C, Razuvaev V, Nieplova E, Cegnar
		T, Antonio~López J, Dahlström B, Moberg A, Kirchhofer W, Ceylan A,
		Pachaliuk O, Alexander L~V and Petrovic P 2002 {\em Int. J. Climatol.\/} {\bf
			22} 1441--1453
		\urlprefix\url{https://rmets.onlinelibrary.wiley.com/doi/abs/10.1002/joc.773}
		
		\bibitem{Besselaar2015}
		Besselaar E~J~M~V~D, Tank A~M~G~K, Schrier G~V~D, Abass M~S, Baddour O, Engelen
		A~F~V, Freire A, Hechler P, Laksono B~I, Iqbal, Jilderda R, Foamouhoue A~K,
		Kattenberg A, Leander R, Güingla R~M, Mhanda A~S, Nieto J~J, Sunaryo,
		Suwondo A, Swarinoto Y~S and Verver G 2015 {\em Bull Am Meteorol Soc\/} {\bf
			96} 16--21 ISSN 00030007, 15200477
		\urlprefix\url{http://www.jstor.org/stable/26219458}
		
		\bibitem{Bunde2017}
		Bunde A and Ludescher J 2017 {\em Long-Term Memory in Climate: Detection,
			Extreme Events, and Significance of Trends\/} (Cambridge University Press) p
		318–339
		
		\bibitem{Massah2016}
		Massah M and Kantz H 2016 {\em Geophys. Res. Lett.\/} {\bf 43} 9243--9249
		\urlprefix\url{https://agupubs.onlinelibrary.wiley.com/doi/abs/10.1002/2016GL069555}
		
		\bibitem{Janosi2002}
		Kir\'aly A and J\'anosi I~M 2002 {\em Phys. Rev. E\/} {\bf 65}(5) 051102
		\urlprefix\url{https://link.aps.org/doi/10.1103/PhysRevE.65.051102}
		
		\bibitem{Kassel2022}
		Kassel J~A and Kantz H 2022 {\em Phys. Rev. Research\/} {\bf 4}(1) 013206
		\urlprefix\url{https://link.aps.org/doi/10.1103/PhysRevResearch.4.013206}
		
		\bibitem{Desmukh2021}
		Deshmukh V, Bradley E, Garland J and Meiss J~D 2021 {\em Chaos: An
			Interdisciplinary Journal of Nonlinear Science\/} {\bf 31} 123102
		\urlprefix\url{https://doi.org/10.1063/5.0069365}
		
		\bibitem{Phillips2023}
		Phillips E~T, H\"oll M, Kantz H and Zhou Y 2023 {\em Phys. Rev. E\/} {\bf
			108}(3) 034301
		\urlprefix\url{https://link.aps.org/doi/10.1103/PhysRevE.108.034301}
		
		\bibitem{Polotzek2021}
		Polotzek K 2021 {\em Modeling and analysis of non-Gaussian long-range
			correlated data-Extreme value theory and effective sample sizes with
			applications to precipitation data\/} Ph.D. thesis Technische Universit{\"a}t
		Dresden Dresden
		
		\bibitem{Massah2023}
		Massah M, Aghion E, Kassel J~A and Kantz H 2023  In preparation
		
	\end{thebibliography}
	
	\providecommand{\newblock}{}

	\newpage
	\appendix
	
	\section{Convexity in either drift or diffusion parameters \label{app:convexity}}
	In this section, we provide more details regarding the uniqueness of the optimal solution of \Eqref{eq:log_likelihood}, showing that the optimisation of $\cS$ in either $f$ or $g$ is unique.
	Since $\cS$ is not convex in $g$, we apply a trick and show that $\cS$ is instead convex in $f$ and
	\begin{equation}
	\psi(x) = \frac{1}{g(x)}.
	\end{equation}
	As 
	\begin{equation}
	\frac{\delta S}{\delta g} = \underbrace{\frac{\delta \psi }{\delta g}}_{\neq 0} \frac{\delta S}{\delta \psi }
	\end{equation}
	shows, it is then sufficient to study the optima in $\psi$, and
	as we assume throughout the work that $g(x) > 0$, this unique solution in $\psi$ defines a unique solution for $g$ too.
	
	The action to be considered now reads
	\begin{equation}
	\fl    \cS[f, \psi] = \frac{\Delta t^2}{2} \sum_{m,n} \psi(x_{m-1}) (v_m - f(x_{m-1})) C^{-1}_{m,n}  (v_n - f(x_{n-1})) \psi(x_{n-1}) - \sum_n \ln |\psi(x_{n-1})|. \quad
	\end{equation}
	We begin by allowing for a general parametrisation of $f(x)$ and $\psi(x)$ by choosing a general set of basis functions $\chi_{p}(x)$ for $p=1,...,P$, $q=1,...,Q$ to write
	\begin{equation}
	\label{appeq:fg_decomposition}
	f(x) = \sum_{p=1}^P f_p \chi_{p}(x) \qquad  \psi(x) = \sum_{q=1}^{Q} \psi_q \chi_{q}(x).
	\end{equation}
	Suitable choices for these basis functions include polynomials $(\chi_p(x) = x^{p})$ or indicator functions of disjoint intervals. The only requirement is that the decomposition in \Eqref{appeq:fg_decomposition} be unique in the coefficients $\{f_p\}, \{\psi_q\}$. The optimisation of the log-likelihood (\Eqref{eq:log_likelihood}) then takes place in these coefficients,
	\begin{eqnarray}
	\fl    \cS[\{f_p\}, \{\psi_q\}] = \\
	\fl    \frac{\Delta t^2}{2} 
	\sum_{m,n=1}^N 
	\left(\sum_{q} \psi_{q} \chi_{q}(x_{m-1}) \right) \left( v_m - \sum_p f_p \chi_p(x_{m-1})\right)
	C^{-1}_{m,n} 
	\left(v_n - \sum_{p'} f_{p'} \chi_{p'}(x_{n-1})\right)\left(\sum_{q'} \psi_{q'}\chi_{q'}(x_{n-1}) \right) \nonumber
	\\
	- \sum_{n=0}^{N-1} \ln \left| \sum_{q} \psi_q \chi_q(x_{n})\right|,
	\end{eqnarray}
	and therefore is a finite-dimensional optimisation problem. 
	
	It remains to show that $\cS$ is convex in either $\{f_p\}$, or $\{\psi_q\}$. We treat $\cS$ as a sum of a bilinear and a logarithmic part, defining $\cS_{\rm log}= - \sum_{n=0}^{N-1} \ln \left| \sum_{q} \psi_q \chi_q(x_{n})\right|$. 
	
	It follows that the Hessian of $\cS_{\rm log}$ is 
	\begin{equation}
	[\partial^2 \cS_{\rm log}]_{p,q}= \frac{\partial^2}{\partial \psi_p \psi_q} \left(- \sum_{n=0}^{N-1} \ln \left| \sum_{p'} \psi_{p'} \chi_{p'}(x_{n})\right|\right) = \sum_n \frac{\chi_p(x_n) \chi_q(x_n)}{\left( \sum_{p'} \psi_{p'} \chi_{p'}(x_{n}) \right)^2}. \quad
	\end{equation}
	For an arbitrary vector $w_{1},...,w_q$, one therefore finds
	\begin{equation}
	\sum_{p,q} w_p  [\partial^2 \cS_{\rm log}]_{p,q} w_q = \sum_n \frac{ \left(\sum_p \chi_p(x_n) w_p \right)^2}{\left( \sum_{p'} \psi_{p'} \chi_{p'}(x_{n}) \right)^2} > 0,
	\end{equation}
	and hence the Hessian is positive definite and $\cS_{\rm \log}$ is convex. 
	
	We next consider  the bilinar contribution to $\cS$ and study the $P \times P$  (or $Q \times Q$)-dimensional diagonal sub-blocks of the Hessian matrix of $\cS_{\rm bil}$ in either $\{f_p\}$ or $\{\psi_q\}$.
	We obtain
	\begin{eqnarray}
	\fl    \frac{\partial \cS}{\partial f_p \partial f_q } = \Delta t^2\sum_{m,n} \chi_p(x_{m-1})  \left(\sum_{p'} \psi_{p'}\chi_{p'}(x_{m-1}) \right) C^{-1}_{m,n}  \left(\sum_{q'} \psi_{q'} \chi{_q'}(x_{n-1}) \right) \chi_q(x_{n-1}), \\
	\fl     \frac{\partial \cS}{\partial \psi_p \partial \psi_q } = \Delta t^2 \sum_{m,n} \chi_p(x_{m-1})  \left[ \left(v_m - \sum_p f_p \chi_p(x_{m-1}) \right) C^{-1}_{m,n}  
	\left(v_n - \sum_q f_q \chi_q(x_{n-1})\right) \right]\chi_q(x_{n-1}). \qquad \quad 
	\end{eqnarray}
	We now consider an arbitrary vector $w_{k}$ with $k=1,...,P$ for drift and $k=1,...,Q$ for diffusion and evaluate
	\begin{equation}
	\sum_{p,q=1}^{P} w_p \frac{\partial \cS}{\partial f_p \partial f_q } w_q \quad \mathrm{and} \quad  
	\sum_{p,q=1}^{Q} w_{p} \frac{\partial \cS}{\partial \psi_p \partial \psi_q } w_{q} \,. 
	\end{equation}
	The first term corresponds to
	\begin{eqnarray}
	\fl    \sum_{p,q=1}^{P} w_p \frac{\partial \cS}{\partial f_p \partial f_q } w_q = \\
	\fl\Delta t^2\sum_{m,n} \left( \sum_p w_p \chi_p(x_{m-1}) \right) \left(\sum_{p'} \psi_{p'}\chi_{p'}(x_{m-1}) \right) C^{-1}_{m,n}  \left(\sum_{q'} \psi_{q'} \chi{_q'}(x_{n-1}) \right) \left( \sum_q w_q \chi_q(x_{n-1}) \right). \qquad\quad
	\end{eqnarray}
	Introducing
	\begin{eqnarray}
	W^-_n &= \sum_p w_p \chi_p(x_{n-1})\\
	\Psi_n &= \sum_{q'} \psi_{q'} \chi{_q'}(x_{n-1})  = \psi(x_{n-1}),
	\end{eqnarray}
	this reads
	\begin{equation}
	\sum_{p,q=1}^{P} w_p \frac{\partial \cS}{\partial f_p \partial f_q } w_q = \frac{\Delta t^2}{2}\sum_{m,n} W_m^- \Psi_m C^{-1}_{m,n} 
	\Psi_n W^-_n.
	\end{equation}
	Since $C^{-1}$ is the inverse of a Gaussian correlation matrix it is a positive definite matrix, and therefore
	\begin{equation}
	\sum_{p,q=1}^{P} w_p \frac{\partial \cS}{\partial f_p \partial f_q } w_q > 0
	\end{equation}
	for all choices of $w$. Hence $\cS$ is convex in $f_p$.
	
	To study convexity in $\psi$, we evaluate the  second term which is
	\begin{eqnarray}
	\fl \sum_{q,q'=1}^{Q} w_{q} \frac{\partial \cS}{\partial \psi_q \partial \psi_{q'} } w_{q'} =\\
	\fl    \Delta t^2 \sum_{m,n} 
	\left( \sum_q w_{q} \chi_q(x_{m-1}) \right)  
	\left(v_m - \sum_p f_p \chi_p(x_{m-1}) \right)
	C^{-1}_{m,n}  
	\left(v_n - \sum_{p'} f_{p'} \chi_{p'}(x_{n-1})\right)
	\left( \sum_{q'} w_{q'} \chi_{q'}(x_{n-1}) \right).\nonumber\\
	\end{eqnarray}
	Further introducing
	\begin{eqnarray}
	W^+_n &= \sum_q w_{q} \chi_q(x_{n-1})\\
	V_n &= v_n - \sum_p f_p \chi_p(x_{n-1}),
	\end{eqnarray}
	this reads
	\begin{equation}
	\sum_{p,q=1}^{Q} w_{p} \frac{\partial \cS}{\partial \psi_p \partial \psi_q } w_{q}
	=
	\Delta t^2 \sum_{m,n} 
	W^+_m 
	V_m
	C^{-1}_{m,n}  
	W^+_n
	V_n.
	\end{equation}
	Again, positive definiteness of $C^{-1}$ implies that
	\begin{equation}
	\sum_{p,q=1}^{Q} w_{p} \frac{\partial \cS}{\partial \psi_p \partial \psi_q } w_{q} > 0
	\end{equation}
	for all choices of $w$ and hence $\cS$ is convex in $\Psi$.
	
	Since both Hessians in $\{f_p\}, \{\psi_p\}$ of the log-likelihood are positiv definite, the log-likelihood is convex in both subspaces and possesses a global minimum in each of the subspaces.

	\section{Optimal drift estimate for fixed diffusion}\label{append:fixed_diffusion}
	We provide further details on the analytic solution given for the minimum solution $\delta S/ \delta \hat{f} \equiv 0$ when $g(x)$ is fixed, see \Eqref{eq:log_likelihood_f}.
	
	Setting out from Equations~(\ref{eq:emp_noise}, \ref{eq:log_likelihood}), the full action reads
	\begin{equation}
	\fl     \cS[f,g|\x] = \frac{1}{2} \sum_{m,n}\frac{\left(f(x_{m-1})\Delta t - (x_m - x_{m-1}) \right)}{g(x_m)}C^{-1}_{m,n}\frac{\left(f(x_{n-1})\Delta t - (x_n - x_{n-1}) \right)}{g(x_n)}\ .
	\label{appeq:action_f}
	\end{equation}
	Since we do not vary $g(x)$, we absorb the inhomogeneous diffusivity into the inverse of the fractional correlation matrix (see \Eqref{eq:corr_fgn}) and introduce the effective propagator
	\begin{equation}
	G_{m,n} = \frac{C^{-1}_{m,n}}{g(x_{m-1})g(x_{n-1})}.
	\end{equation}
	Further introducing $v_n = (x_n-x_{n-1})/\Delta t$, one readily recovers the expression given in \Eqref{eq:log_likelihood_f}.
	
	Technically, this action is minimised by the ``empirical force'' $f(x_{n-1}) = v_n$. 
	In order to effectively average over the fluctuations of $x_n$, however, a low-dimensional representation of $f$ is more suitable.
	We choose a polynomial representation, \ie \mbox{$f(x) = \sum_{\ell=0}^{L-1} f_{\ell} x^{\ell}$} with $L \ll N$. 
	Inserting this ansatz into \Eqref{appeq:action_f}, one finds
	\begin{eqnarray}
	\fl     \cS[f,g|\x] &= \frac{(\Delta t)^2}{2} \sum_{\ell,k} \sum_{m,n} \left(f_{\ell}x_{m-1}^{\ell} - v_m \right)G_{m,n}\left(f_{k}x_{n-1}^{k} - v_n \right) \\
	\fl     &= \frac{(\Delta t)^2}{2} \left\lbrace \sum_{\ell,k} f_{\ell} f_{k} \left( \sum_{m,n} x_{m-1}^{\ell} G_{m,n} x_{n-1}^{l}\right) - 2 \sum_{\ell} f_{\ell} \left( \sum_{m,n} x_{m-1}^{\ell} G(g)_{m,n} (x_{n} - x_{n-1}) \right)   \right\rbrace + \ldots, \qquad
	\end{eqnarray}
	where we ignore $f$-independent terms. Identifying
	\begin{eqnarray}
	H_{\ell,k}&=\sum_{m,n} x_{m-1}^{\ell} G(g)_{m,n} x_{n-1}^{k}\\
	J_{\ell} &= \sum_{m,n} x_{m-1}^{\ell} G(g)_{m,n} (x_{n} - x_{n-1}),
	\end{eqnarray}
	the optimal point $\nabla_{\hat{f}} \cS =  0$ implies $(\Delta t)^2 \left\lbrace H \hat{f} -  J\right \rbrace = 0$, and hence $\hat{f} = H^{-1} J$ as stated in the main text.
	
	\section{fOMo estimation for the fractional Ornstein-Uhlenbeck process}\label{append:FOU}
	In the special case of linear drift and constant diffusion, \Eqref{eq:FLE_def} recovers the Euler-Maruyama discretisation of a fractional Ornstein Uhlenbeck process, \ie 
	\begin{equation}
	x_{n+1} - x_{n} = - f  x_n \dt + g \dWh_{n+1}.
	\end{equation}
	Accordingly parametrising $f(x) = f x, g(x) = g$, the fractional Onsager Machlup action (\Eqref{eq:log_likelihood}) reduces to the two-dimensional function
	\begin{equation}
	\cS[f,g] = \frac{(\Delta t)^2}{2 g^2} \sum_{m,n}\left( f - v_m\right)C_{m,n}^{-1} \left( f-v_n \right) + \ln |g|.
	\end{equation}
	The minimum in $f$ is $g$-independent and is given by
	\begin{equation}
	\hat{f} = \frac{\sum_{m,n}C^{-1}_{m,n} v_n}{\sum_{m,n}C_{m,n}^{-1}},
	\end{equation}
	which extends \cite{PerezGarcia2018} from white noise to  processes driven by Gaussian noise with arbitrary correlations.
	The minimum in $g$ is given by
	\begin{equation}
	\hat{g}^2 = (\Delta t)^2 \sum_{m,n}\left( \hat{f} - v_m\right)C_{m,n}^{-1} \left( \hat{f}-v_n \right)\ .
	\end{equation}

	\section{A Note on the Numerics}\label{appendix:numerics}
	In this section, we elaborate on the computation of the stochastic action given by \Eqref{eq:log_likelihood}.
	For stationary processes, the autocovariance function solely depends on the difference between two time instances. Hence, the covariance matrix $C_{ij}$ is a $(N-1)\times (N-1)$ symmetric Toeplitz matrix, where $N$ is the length of the time series, i.e.  ${C_{ij} = (\dt)^{2H} \times \left( \left| i-j +1 \right|^{2H} + \left| i-j - 1 \right|^{2H} - 2 \left|i - j \right|^{2H} \right)} =: C(|i-j|\Delta t)$, in which $i,j\in\{1,\ldots,N-1\}$. We compute the inverse correlation matrix, the propagator, exploiting this fact.
	The inverse of a Toeplitz matrix may be expressed as \cite{Lv2007}
	\begin{equation}
	C^{-1} = T_1 U_1 + T_2 U_2 \,,
	\end{equation} 
	in which $T_1$ and $T_2$ are Toeplitz matrices and $U_1$, $U_2$ are upper triangular matrices with Toeplitz structure:
	\begin{eqnarray}
	T_1 = \left(\begin{array}{cccc}
	y_1 & y_n & \ldots  & y_2 \\
	y_2 & y_1 & \ddots & \\
	\vdots & \ddots & \ddots & y_n \\
	y_n & \ldots & y_2 & y_1
	\end{array}\right) \,, \quad
	U_1 = \left(\begin{array}{cccc}
	1 & -x_n & \ldots & -x_2 \\
	& 1 & \ddots & \vdots \\
	& & \ddots & -x_n \\
	& & & 1
	\end{array}\right) \\
	T_2 = \left(\begin{array}{cccc}
	x_1 & x_n & \ldots & x_2 \\
	x_2 & x_1 & \ddots & \vdots \\
	\vdots & \ddots & \vdots & x_n \\
	x_n & \ldots & x_2 & x_1
	\end{array}\right) \,, \quad
	U_2 = \left(\begin{array}{cccc}
	0 & y_n & \ldots & y_2\\
	& 0 & \ddots & \vdots \\
	& & \ddots & y_n \\
	& & & 0
	\end{array}\right) \,.
	\end{eqnarray}
	The vectors $x$ and $y$ are solutions to the linear equations
	\begin{eqnarray}
	&C x = h \label{eq:action_numerics_1} \,,\\
	&C y = e_1 \label{eq:action_numerics_2} \,,\\
	e_1 = \left(\begin{array}{c}
	1 \\ 0\\ \vdots \\ 0
	\end{array}\right) \,, &\quad
	h = \left(\begin{array}{c}
	0 \\ C((N-1)\Delta t) - C(\Delta t) \\ \vdots \\
	C(2\Delta t) - C((N-2)\Delta t) \\ C(\Delta t) - C((N-1)\Delta t)
	\end{array}\right) \,.
	\end{eqnarray}
	The matrices $T$ and $U$ are entirely determined by $x$ and $y$. Hence, only these two vectors have to be saved during computation. We obtain $x$ and $y$ solving the linear equations \Eqref{eq:action_numerics_1} and \Eqref{eq:action_numerics_2} using the Levinson-Durbin algorithm, which has complexity $\mathcal{O}(N^2)$.
	Thus, the action reads
	\begin{eqnarray}
	\fl    \cS[\vec{x}|f,g] &=  \frac{(\Delta t)^2}{2} \sum_{m,n=1}^N \frac{\left(f_{m-1} - v_m \right)}{g_{m-1}} \, (T_1 U_1 + T_2 U_2)_{mn} \, \frac{\left(f_{n-1} - v_n \right)}{g_{n-1}}
	+ \sum_{n=0}^{N-1} \ln |g_n| \qquad 
	\end{eqnarray}
	which are repeated Toeplitz matrix multiplications. During the optimisation of $f$ and $g$, $C^{-1}$ does not change. Thus, we only compute Toeplitz matrix multiplications throughout the optimisation.
	
	\section{Finite-Size Error Scaling of Synthetic Data}\label{append:error_scaling}
	We investigate the finite-size error scaling as a function of the trajectory length $N$ of synthetic data generated from the model (see main text)
	\begin{equation}
	x_{n+1} = x_n + \Delta t (-0.25 \, x_n^3 + 0.5 \, x_n) + (0.2 \, x_n^2 + 0.5) \Delta \xi_{n+1}^H \,, \nonumber
	\end{equation}
	with $\Delta t = 0.01$. For fixed Hurst exponent and an ensemble of $100$ trajectories, we fit a power law to the asymptotic scaling of the RMSE in $N$, up to $N=2\times10^6$.
	In accordance with the central limit theorem,  we find $\mathrm{RMSE}\sim 1/\sqrt{N}$ in the case $H=\frac{1}{2}$ for both drift and diffusion. 
	However, we observe that the inference error of the drift for $H\neq\frac{1}{2}$ scales according to $\mathrm{RMSE}\sim N^{\alpha}$ with $\alpha \sim H^2$, while the diffusion error scaling does not show a clear trend. 
	We further observe qualitatively similar error scaling for other toy models, e.g. monostable potentials.
	Scaling behaviour deviating from the central limit theorem for correlated data is also observed for the sample mean of correlated Gaussian random variables. Its standard deviation scales like $\sigma_{\mu_{N}} \sim N^{H-1}$ \cite{Polotzek2021, Massah2023}. Hence, the convergence speed of the sample mean depends on the strengths of the correlations, also allowing faster convergence for $H<\frac{1}{2}$, which we observe as well.
	
	\begin{figure}
		\centering
		\includegraphics[width=0.49\textwidth]{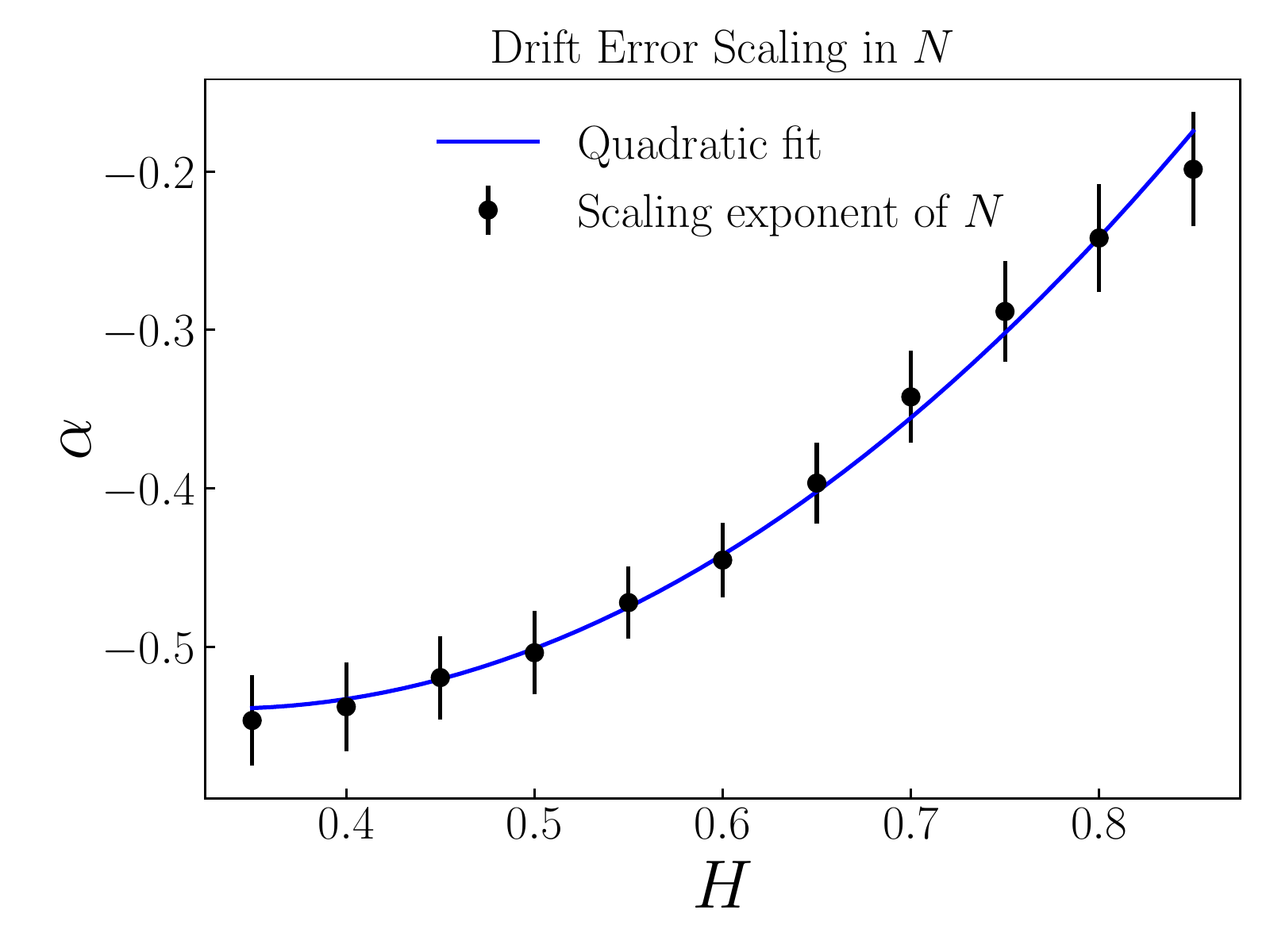}
		\includegraphics[width=0.49\textwidth]{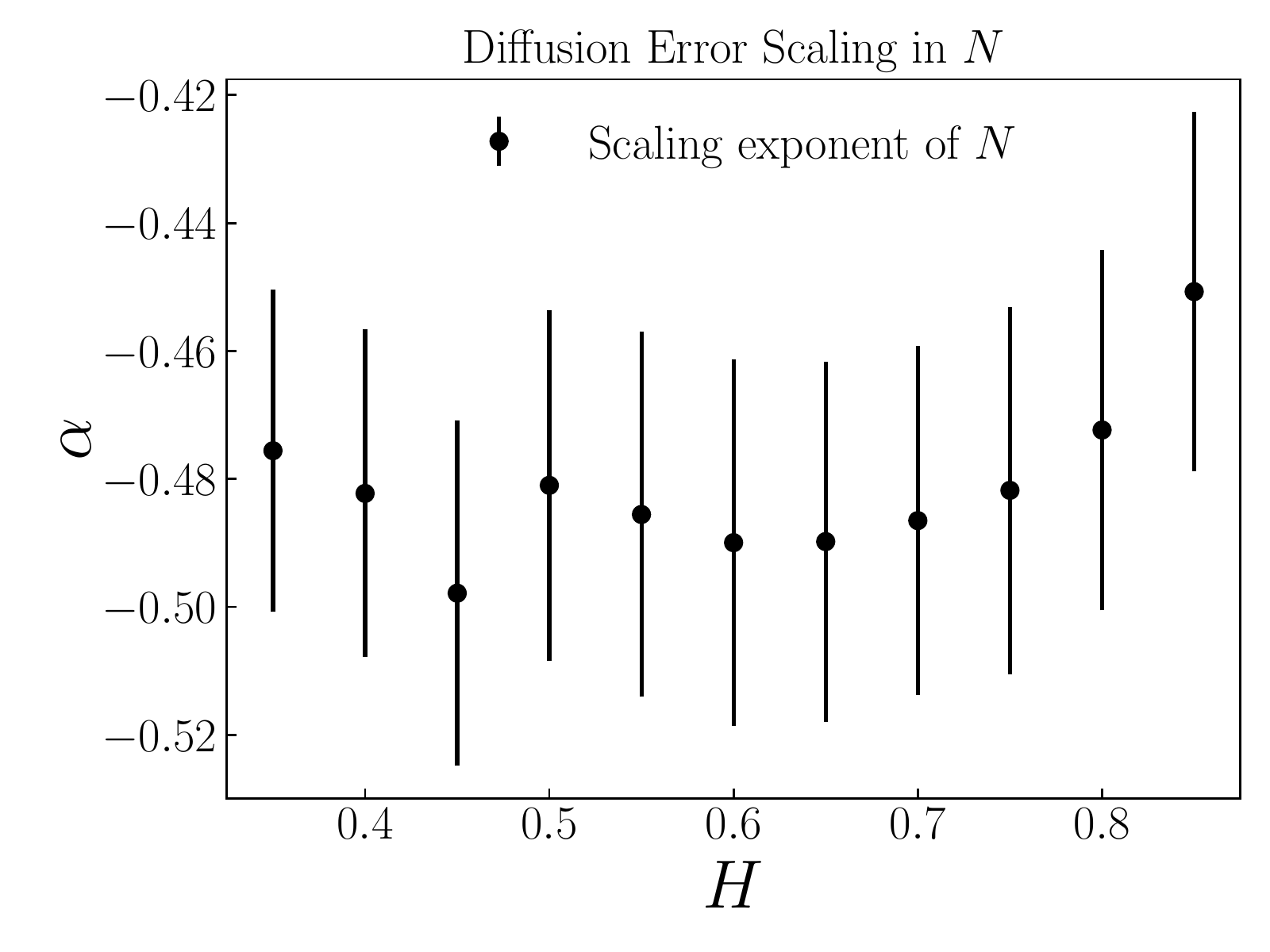}
		\caption{Toy model with double well potential and quadratic diffusion term. Error scaling in trajectory length $N$ as a function of the Hurst exponent $H$.}
		\label{fig:appendix_error_scaling}
	\end{figure}

	\section{Potsdam Temperature Data}\label{append:temperature}
	In the main text, we demonstrate fOMo by reconstructing a stochastic model from daily mean temperature anomalies recorded at Potsdam Telegrafenberg weather station, Potsdam, Germany. Here, we elaborate on the procedure.
	We obtain the temperature anomalies by subtracting the seasonal cycle. In the main text, we state that the resulting time series is approximately stationary. In fact, there are slow-mode variations in the time series as well as a warming trend. 
	However, these only marginally violate the stationarity of the time series.
	We thus abstain from subtracting an additional trend from the data besides  seasonality. 
	Furthermore, we neglect measurement errors in our analysis since they are sufficiently small compared to the dynamics.

\end{document}